\documentclass[12pt]{article}
\usepackage{amsmath,amssymb,amsfonts}
\usepackage{graphicx,epsfig}
\usepackage{color}
\usepackage{slashed}
\begin{document}

\def\simgt{\mathrel{\lower2.5pt\vbox{\lineskip=0pt\baselineskip=0pt
           \hbox{$>$}\hbox{$\sim$}}}}
\def\simlt{\mathrel{\lower2.5pt\vbox{\lineskip=0pt\baselineskip=0pt
           \hbox{$<$}\hbox{$\sim$}}}}

\begin{titlepage}

\begin{flushright}
UCB-PTH-06/14\\
LBNL-61281
\end{flushright}

\vskip 1.5cm

\begin{center}
{\Large \bf Supersymmetry without a Light Higgs Boson}

\vskip 1.0cm

{\large Riccardo Barbieri$^a$, Lawrence J.~Hall$^{b,c}$, Yasunori Nomura$^{b,c}$, \\
  Vyacheslav S.~Rychkov$^a$}\\[1cm]

{\it $^a$ Scuola Normale Superiore and INFN, Piazza dei Cavalieri 7, 
  I-56126 Pisa, Italy}\\
{\it $^b$ Department of Physics, University of California, 
  Berkeley, CA 94720, USA}\\
{\it $^c$ Theoretical Physics Group, Lawrence Berkeley National Laboratory, 
  Berkeley, CA 94720, USA}\\[4mm]

\vskip 1.0cm

\abstract{Motivated by the absence, so far, of any direct signal of 
conventional low-energy supersymmetry, we explore the consequences of 
making the lightest Higgs boson in supersymmetry relatively heavy, up 
to about $300~{\rm GeV}$, in the most straightforward way, i.e. via 
the introduction of a chiral singlet $S$ with a superpotential interaction 
with the Higgs doublets, $\lambda S H_1 H_2$.  The coupling $\lambda$ 
dominates over all the other couplings and, to maintain the successful 
perturbative analysis of the ElectroWeak Precision Tests, is only 
restricted to remain perturbative up to about $10~{\rm TeV}$.  The 
general features of this ``$\lambda$SUSY'' framework, which deviates 
significantly from the MSSM or the standard NMSSM, are analyzed in 
different areas: ElectroWeak Precision Tests, Dark Matter, naturalness 
bounds on superparticle masses, and LHC signals.  There is a rich 
Higgs/Higgsino sector in the (200~--~700)~GeV mass region, which may 
include LSP Higgsino dark matter.  All other superpartners, apart 
from the top squarks, may naturally be heavier than 1--2~TeV.  This 
picture can be made consistent with gauge coupling unification.}

\end{center}

\end{titlepage}

\section{Introduction}

The naturalness problem of the Fermi scale in the Standard Model (SM) amounts 
to understand the lightness of the Higgs boson relative to any mass scale in 
the theory, whatever it may be, that completes the SM itself at high energies, 
or in the Ultra-Violet (UV).  If the relative Higgs lightness is not accidental, 
this gives the best hope for seeing new physics at the LHC, hence the focus 
on the naturalness problem.  As is well known, the constraints on a positive 
solution to this problem largely come from the success of the SM in accounting 
for the results of the ElectroWeak Precision Tests (EWPT).  In fact, the 
difficulty to solve it is exacerbated by the apparent lightness of the Higgs 
boson, as indirectly implied by the EWPT~\cite{unknown:2005em}.  If, for one 
reason or another, we were misled in interpreting the data of the EWPT in 
terms of a light Higgs mass and the Higgs boson were heavier, the upper bound 
on the naturalness cutoff of the theory of the electroweak interactions might 
be relaxed by a non-negligible amount~\cite{Barbieri:2005kf,Barbieri:2006dq}.

This simple observation has a counterpart in supersymmetry, which 
is in many respects the most natural UV completion of the Standard 
Model (SM). In supersymmetric extensions of the SM, like the Minimal 
Supersymmetric Standard Model (MSSM), there is no problem in letting the 
cutoff become arbitrarily high, while the lightest Higgs boson is actually 
predicted to be light.  So light in fact that to make its mass consistent 
with the experimental lower bound (about $114~{\rm GeV}$ in most of the 
parameter space~\cite{Barate:2003sz}) requires a large radiative correction 
due to a heavy stop, which in turn brings back at least a few percent 
fine-tuning in the $Z$ boson mass.  Accommodating a heavier Higgs boson 
would therefore ease the naturalness problem in supersymmetry too, 
while allowing heavier superpartners.  This has motivated several 
attempts in the literature to increase the lightest Higgs boson 
mass~[\ref{Espinosa:1998re:X}~--~\ref{Gripaios:2006nn:X}].

Here we explore the possibility of making the lightest Higgs boson in 
supersymmetry relatively heavy in the most straightforward way, i.e. via the 
introduction of a chiral singlet $S$ with a large superpotential interaction 
$\lambda S H_{1} H_{2}$, where $H_{i}$ are the two Higgs doublet multiplets, 
leading to a large quartic interaction of the Higgs doublets.  Unlike 
the usual treatment of the Next to Minimal Supersymmetric Standard Model 
(NMSSM)~\cite{Bastero-Gil:2000bw}, $\lambda$ is not perturbative up to 
the unification scale.  We insist only that consistency with the EWPT can 
be analyzed in a fully perturbative manner, which we ensure by requiring 
that $\lambda$ remains perturbative up to $\approx 10~{\rm TeV}$.  At this 
scale a change of regime of the theory should intervene, which we leave 
unspecified.  It is an open question whether this will allow a successful 
unification of the gauge couplings.  A positive existence proof of this 
is given in Refs.~\cite{Harnik:2003rs,Chang:2004db,Birkedal:2004zx}.

With perturbativity preserved up to $\approx 10~{\rm TeV}$, contributions 
to the EWPT from unknown UV physics can be sufficiently small.  There 
may, however, still be calculable ``infrared'' contributions which make 
the theory incompatible with the EWPT.  In particular, one might think that 
contributions from a lightest Higgs boson of mass $(200\!\sim\!300)~{\rm GeV}$, 
as well as those from the other Higgs sector particles which grows like the 
fourth power of $\lambda$, are dangerous for the EWPT.  Quite on the contrary, 
our general analysis of the contributions from the Higgs boson sector and, 
even more importantly from the Higgsino sector, shows that the theory is 
perfectly consistent with the EWPT in a significant range of the parameter 
space.  Deviations from the ``EWPT ellipse'' due to a relatively heavy 
lightest Higgs boson are entirely canceled by the contributions from the 
other Higgs bosons and the Higgsinos, in almost all the parameter space 
in which electroweak symmetry breaking occurs naturally.

In Section~\ref{sec:lambdasusy} we define our theory, minimize the scalar 
potential for large $\lambda$, and give both the scalar and fermion spectrum. 
Many significant features of the theory are independent of the specific 
forms for the superpotential and soft operators, but depend critically on 
$\lambda$ being large.  Hence we call this framework ``$\lambda$SUSY.'' 
In Section~\ref{sec:EWPT} we consider the EWPT from both scalar and fermion 
sectors.  In Section~\ref{sec:DM} we consider the candidate for Dark Matter 
(DM) in the special case of heavy weak gauginos, as allowed by naturalness. 
In Section~\ref{sec:NC} we obtain the naturalness bounds on the various 
particles in the limiting case in which the scalar $S$ is taken heavy. 
The characteristic manifestations of the model at the LHC are described 
in Section~\ref{sec:LHC}.  Summary and Conclusions are given in 
Section~\ref{sec:OC}.

\section{The {\boldmath $\lambda$}SUSY Model}
\label{sec:lambdasusy}

We consider the most general supersymmetric theory with $SU(3)_C \times SU(2)_L 
\times U(1)_Y$ gauge invariance that contains a singlet chiral superfield $S$ 
in addition to the fields of the MSSM.  The theory possesses the superpotential 
interaction
\begin{equation}
  W_{\lambda} = \lambda\, S H_{1} H_{2},
\label{eq:lambda}
\end{equation}
where $H_{1}$ and $H_{2}$ are the usual Higgs doublets, coupled respectively 
to the down-type and up-type quarks.  We increase the mass of the lightest 
Higgs boson, and therefore the naturalness of the theory, by taking $\lambda$ 
large.  We call this the $\lambda$SUSY model, to stress the key role of the 
large coupling $\lambda$.  Such a large coupling may arise from compositeness 
of (some of) the Higgs states~\cite{Harnik:2003rs,Chang:2004db,Birkedal:2004zx,%
Delgado:2005fq}.

How large can $\lambda$ be?  To keep perturbativity up to at least 
$10~{\rm TeV}$, we require $\lambda^{2}(\Lambda=10~{\rm TeV})$ be less 
than about $4\pi$. Given the Renormalization Group Equation (RGE) satisfied 
by $\lambda$ ($t=\ln{E}$)
\begin{equation}
  \frac{d\lambda^{2}}{dt} = \frac{\lambda^{4}}{2\pi^{2}},
\end{equation}
this implies that, at the low-energy scale of about $500~{\rm GeV}$, $\lambda$ 
is less than about $2$. Throughout the paper, unless differently stated, we take 
$\lambda(500~{\rm GeV})=2$.  With this value of $\lambda$ the Landau pole is 
typically a few tens of TeV.

The superpotential of our model takes the general form
\begin{equation}
  W = \mu(S)H_{1}H_{2} + f(S),
\label{eq:superpot}
\end{equation}
and, neglecting the gauge terms, the scalar potential can be written in 
the form
\begin{equation}
  V = \mu_{1}^{2}(S) |H_{1}|^{2} + \mu_{2}^{2}(S) |H_{2}|^{2}
    - (\mu_{3}^{2}(S) H_{1} H_{2} + {\rm h.c.})
    + \lambda^{2} |H_{1}H_{2}|^{2} + V(S).
\label{eq:pot}
\end{equation}
The mass parameters $\mu_{1}^{2}(s)$, $\mu_{2}^{2}(s)$, $\mu_{3}^{2}(s)$, 
$\mu(s)$ and
\begin{equation}
  M(s) \equiv \frac{d^{2}f(s)}{ds^{2}},
\end{equation}
calculated at the background expectation value $s$ of the field $S$, 
characterize many of the properties of the theory.  We generally denote 
them as $\mu_{1}^{2}$, $\mu_{2}^{2}$, $\mu_{3}^{2}$, $\mu$, and $M$, by 
leaving understood their argument $s$.  For simplicity we assume $CP$ 
invariance of $V$ and $W$.  This formulation of the $\lambda$SUSY theory 
is convenient because many equations take a similar form to the familiar 
MSSM case.

The stability of the potential (\ref{eq:pot}) requires
\begin{equation}
  \mu_{1}^{2},\, \mu_{2}^{2} > |\mu|^2,
\label{eq:stab}
\end{equation}
for all the values of $s$, whereas the condition for electroweak symmetry 
breaking is
\begin{equation}
  |\mu_{3}^{2}| > \mu_{1}\mu_{2},
\end{equation}
at the minimum of the effective potential for $s$, obtained by replacing 
$H_1$ and $H_2$ by their $s$-dependent expectation values.  Under the condition 
(\ref{eq:stab}) electromagnetism is unbroken.  To be more precise, once the 
gauge contribution to the potential is introduced, the potential is always 
stable, and the condition (\ref{eq:stab}), with an additional small term 
on the right-hand-side, becomes the condition for unbroken electromagnetism. 
For our purposes, this difference has negligible impact.  In general we neglect 
in this section all the relatively small effects of the $SU(2)_L \times U(1)_Y$ 
$D$-term contributions.  We shall also neglect the possibility of spontaneous 
$CP$ violation, i.e. we shall consider all the mass parameters real.%
\footnote{A physical phase in the Higgsino mass matrix (see below) 
could have relevant consequences.}

By minimizing the potential it is straightforward to find
\begin{align}
  \tan\beta & \equiv\frac{v_{2}}{v_{1}} = \frac{\mu_{1}}{\mu_{2}},\\
  \lambda^{2}v^{2} & = \frac{2\mu_{3}^{2}}{\sin2\beta}-\mu_{1}^{2}-\mu_{2}^{2},
\label{eq:lambdav}
\end{align}
where $\lambda = \partial \mu(S)/\partial S$, and $v_{1}$, $v_{2}$, 
$v \equiv (v_1^2 + v_2^2)^{1/2}$ are the usual Vacuum Expectation Values 
(VEVs) of the Higgs fields.  One also finds that the mass of the charged 
Higgs bosons, $H^{\pm}$, is
\begin{equation}
  m_{H^{\pm}}^{2} = \mu_{1}^{2} + \mu_{2}^{2},
\label{eq:mHpm}
\end{equation}
where $\mu_{1}^{2}$ and $\mu_{2}^{2}$ should be evaluated at the minimum of $s$. 
Unlike the mass of $H^{\pm}$, the masses of the neutral scalars $h,H$ and $A$ 
can be expressed in terms of $\mu_{1},\mu_{2}$ and $\mu_{3}$ only if their mixing 
with the scalar $S$ can be neglected.  In this case, the mass of the pseudoscalar 
Higgs state is given by
\begin{equation}
  m_{A}^{2} = \frac{2\mu_{3}^{2}}{\sin2\beta},
\label{eq:mA}
\end{equation}
while the two $CP$-even Higgs states $h_i \equiv \sqrt{2}({\rm Re}H_i^0 - v_i)$ 
mix with the following mass matrix:
\begin{equation}
  \left( \begin{array}[c]{cc}
    m_{A}^{2}\sin^2\!\beta & (\lambda^{2}v^{2}-\frac{1}{2}m_{A}^{2})\sin2\beta \\
    (\lambda^{2}v^{2}-\frac{1}{2}m_{A}^{2})\sin2\beta & m_{A}^{2}\cos^2\!\beta
  \end{array} \right),
\end{equation}
where $\mu_3^2$ in Eq.~(\ref{eq:mA}) should be evaluated at the minimum of $s$. 
The masses and composition of the mass eigenstates $H$ and $h$ are given by
\begin{equation}
  m_{H,h}^{2} = \frac{1}{2}(m_{A}^{2} 
    \pm X),
\qquad
  X^{2} = m_{A}^{4} - 4\lambda^{2}v^{2}m_{H^{\pm}}^{2}\sin^{2}2\beta,
\end{equation}
and
\begin{eqnarray}
  H &=&  \cos\alpha\,h_{1} + \sin\alpha\,h_{2},
\\
  h &=& -\sin\alpha\,h_{1} + \cos\alpha\,h_{2},
\end{eqnarray}
respectively, where 
\begin{equation}
  \tan\alpha = \frac{m_{A}^{2}\cos2\beta+X}
    {(\lambda^{2}v^{2}-m_{H^{\pm}}^{2})\sin2\beta}.
\end{equation}
We find that the following inequalities hold:
\begin{equation}
  m_{h} \leq \sin\beta\, \lambda v,
\qquad
  m_{H^{\pm}} \leq m_{H} < m_{A}.
\end{equation}

In the fermion sector, the $SU(2)_L \times U(1)_Y$ gauginos are mixed with 
the other fermions only by the relatively small gauge terms.  Since there is 
also no significant naturalness upper bound on their masses, it is relevant 
to consider the case in which these gauginos are decoupled from the other 
fermions.  In this case the charged Higgsino has a mass $\mu$ and the mass 
matrix of the neutral fermions
\begin{equation}
  N_{1} = \frac{1}{\sqrt{2}}(\tilde{H_{1}}-\tilde{H_{2}}),
\quad
  N_{2} = \frac{1}{\sqrt{2}}(\tilde{H_{1}}+\tilde{H_{2}}),
\quad
  N_{3} = \tilde{S},
\end{equation}
is given in terms of $\mu$ and $M$ by
\begin{equation}
  \mathcal{M} = \left(\begin{array}[c]{ccc}
    \mu & 0 & \lambda\frac{v_{1}-v_{2}}{\sqrt{2}}\\
    0 & -\mu & -\lambda\frac{v_{1}+v_{2}}{\sqrt{2}}\\
    \lambda\frac{v_{1}-v_{2}}{\sqrt{2}} & -\lambda\frac{v_{1}+v_{2}}{\sqrt{2}} & M
  \end{array} \right),
\label{eq:ferm_mass}
\end{equation}
where $\mu$ and $M$ are, again, evaluated at the minimum of $s$.  Only 
the relative sign of $\mu$ and $M$ is physical, and we fix our convention 
to be $\mu>0$ for definiteness.  Note that the condition (\ref{eq:stab}) 
implies the inequality
\begin{equation}
  \mu \leq \cos{\beta}\,m_{H^{\pm}}.
\label{eq:mu-bound}
\end{equation}

The $N_{i}$'s will be related to the mass eigenstates (neutralinos) $\chi_{i}$ 
by a non-trivial $3 \times 3$ mixing matrix $V$:
\begin{equation}
  N_{i} = V_{ij} \chi_{j},
\quad
  V^{T}V = 1,
\quad
  V^{T} \mathcal{M} V = \text{diag}(m_{1},m_{2},m_{3}).
\end{equation}
The lightest neutralino $\chi$ has a mass satisfying
\begin{equation}
  |m_{\chi}| \leq \mu,
\end{equation}
and the other two eigenvalues are above $\mu$ and below $-\mu$, respectively. 
Moreover, the determinant of the mass matrix, and thus $m_{\chi},$ vanishes 
for a particular value of $M$:
\begin{equation}
  m_{\chi} = 0 
\qquad \text{for} \quad
  M = -\frac{\lambda^{2}v^{2}}{\mu} \sin2\beta.
\label{eq:light-chi}
\end{equation}
Thus there is a region of parameter space with a naturally light neutralino, 
which is an admixture of $\tilde{S}$ and the Higgsinos.  In fact, the lightest 
neutralino is much lighter than the characteristic scales for $\mu$ and $M$ 
in a wide region around the point satisfying Eq.~(\ref{eq:light-chi}).

The mixing matrix and the spectrum are difficult to determine analytically, 
except in the limiting case of $\tan\beta = 1$, which leads to
\begin{align}
  m_{1} &= \mu,
\\
  m_{2,3} &= \frac{1}{2}(M- \mu \pm Y),
\qquad
  Y^{2} = (M+\mu)^{2}+4\lambda^{2}v^{2}.
\end{align}
In this case $\chi_{1} = N_{1}$, while $\chi_{2,3}$ are mixtures of $N_{2,3}$ 
with a mixing angle $\gamma$, where $\tan\gamma = -(M+\mu+Y)/2\lambda v$. 
The lightest state is either $\chi_{1}$ or $\chi_{2}$ depending on whether 
$M$ is above or below a critical value $M_{c} = \mu - \lambda^{2}v^{2}/2\mu$. 

For sufficiently small $\tan\beta-1$, one can approximately diagonalize 
the matrix (\ref{eq:ferm_mass}) expanding in $\mathcal{M}_{13}$.  The mixing 
between $N_{1}$ and $N_{2,3}$ will be small everywhere except near $M=M_{c}$. 
As a result, the coupling of the lightest neutralino $\chi$ to the $Z$ boson 
will be generically suppressed.  This may play an important role in the 
interpretation of DM, as discussed in Section~\ref{sec:DM}.

When $\mu_{1}=\mu_{2}$, giving $\tan{\beta}=1$, an $SU(2)$ custodial symmetry 
survives, and plays an important role.  In this limit, in the scalar sector 
either $h$ or $H$, depending on the sign of $2 m_{H^\pm}^2 - m_A^2$, behaves 
like the SM Higgs boson, of mass $m_{h} = \lambda v/ \sqrt{2}$, whereas the 
other scalar with $H^\pm$ forms a degenerate $SU(2)$-triplet.  Similarly, 
in the fermion sector, the chargino and $N_{1}$ form a degenerate triplet 
of mass $\mu$.

\section{ElectroWeak Precision Tests}
\label{sec:EWPT}

We perform the analysis of the EWPT in the usual $S$-$T$ plane, with the 
experimental contours taken from~\cite{Plot}.  With respect to the SM, 
the $\lambda$SUSY model has: 1) a modified contribution to $S$ and $T$ 
from the scalar Higgs sector~\cite{Harnik:2003rs,Gripaios:2006nn}; 2) new 
and relevant contributions from the stop-sbottom and from the Higgsinos 
(we consider heavy gauginos).  The contributions of 1) and 2) can be 
analyzed separately. 

To locate our model in the $S$-$T$ plane \textit{before the stop-sbottom 
and Higgsino contributions are added}, we subtract the one-loop SM Higgs 
contributions 
\begin{align}
  T_{\text{Higgs}}(m_{h}) &= -3\left[ A(m_{h},m_{W})-A(m_{h},m_{Z})\right],
\label{eq:sm-T}\\
  S_{\text{Higgs}}(m_{h}) &= F(m_{h},m_{Z}) + m_{Z}^2 G(m_{h},m_{Z}),
\label{eq:sm-S}
\end{align}
from the SM values of $S$ and $T$ (see Appendix~\ref{app:A} for the 
definitions), and then add the one-loop 2~Higgs-Doublet Model (2HDM) 
contributions (computed under the assumption of no doublet-singlet 
mixing)~\cite{Chankowski:2000an,Chankowski:1999ta}
\begin{align}
  T_{\text{2HDM}} = 
    & \sin^{2}(\beta-\alpha)\, 
    [T_{\text{Higgs}}(m_{h}) + A(m_{H^{\pm}},m_{H}) - A(m_{A},m_{H})]
\nonumber\\
    & + \cos^{2}(\beta-\alpha)\,
    [T_{\text{Higgs}}(m_{H}) + A(m_{H^{\pm}},m_{h}) - A(m_{A},m_{h})]
\nonumber\\
    & + A(m_{H^{\pm}},m_{A}),
\label{eq:us-T}\\
  S_{\text{2HDM}} = 
    & \sin^{2}(\beta-\alpha)\,
    [S_{\text{Higgs}}(m_{h}) + F(m_{A},m_{H})]
\nonumber\\
    & + \cos^{2}(\beta-\alpha)\, 
    [S_{\text{Higgs}}(m_{H}) + F(m_{A},m_{h})]
\nonumber\\
    & - F(m_{H^{\pm}},m_{H^{\pm}}).
\label{eq:us-S}
\end{align}
Here, $m_h$ in Eqs.~(\ref{eq:sm-T},~\ref{eq:sm-S}) represents a reference 
value of the Higgs boson mass in the SM, while that in Eqs.~(\ref{eq:us-T},%
~\ref{eq:us-S}) is the mass of the lightest $CP$-even Higgs boson in the 
$\lambda$SUSY model.

\begin{figure}[ptb]
\begin{center}
  \includegraphics[width=8cm]{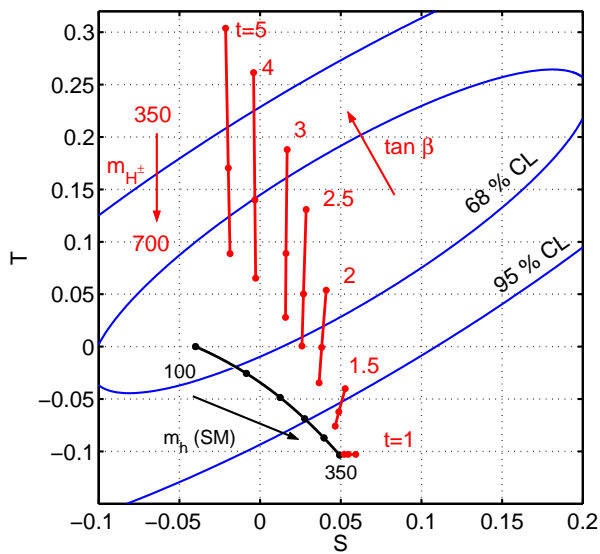}
\end{center}
\caption{The black (darker) curve shows the SM results with a Higgs mass 
 $m_{h} = 100-350~{\rm GeV}$ in $50~{\rm GeV}$ increments.  The ellipses 
 show the regions of the $S$-$T$ plane allowed by EWPT at $1\sigma$ and 
 $2\sigma$.  The red (lighter) curves give the predictions from the Higgs 
 scalar sector in $\lambda$SUSY, as described in the text, with values 
 of $\tan\beta$ in the interval $\tan\beta = 1\!\sim\!5$ as indicated 
 and $m_{H^{\pm}}=350,500,700~{\rm GeV}$.}
\label{fig:EWPT_scalars}
\end{figure}

In the $\lambda$SUSY model, the 6 parameters appearing in Eqs.~(\ref{eq:us-T},%
~\ref{eq:us-S}), i.e. $m_h$, $m_H$, $m_A$, $m_{H^{\pm}}$, $\alpha$ and 
$\beta$, depend only on $\tan\beta$, $m_{H^{\pm}}$ and $\lambda$, which 
we set equal to 2.  In Fig.~\ref{fig:EWPT_scalars} we show the result for 
the EWPT for several values of $\tan\beta$ and $m_{H^{\pm}}$ ($\tan\beta 
= 1\!\sim\!5$ and $m_{H^{\pm}} = 350,500,700~{\rm GeV}$).
Two features are manifest from this figure: i) the role of the custodial 
symmetry for $\tan\beta$ approaching unity, thus suppressing the corrections 
to $T$; ii) the fact that the positive $T$-correction brings most of the 
points of the $\lambda$SUSY model inside the region preferred by experiments, 
at least as long as $\tan{\beta}$ is not too large.

\begin{figure}[ptb]
\begin{center}
  \includegraphics[width=8cm]{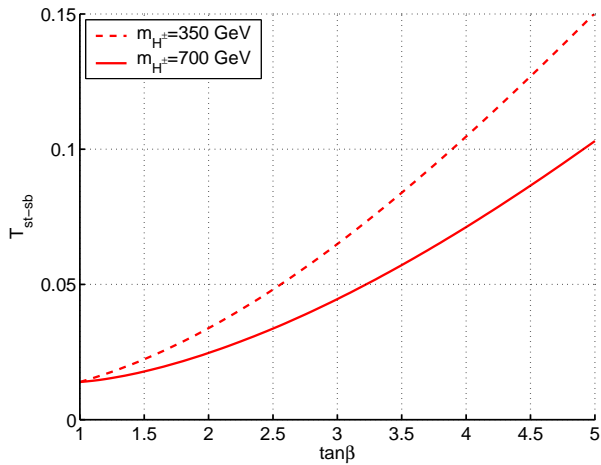}
\end{center}
\caption{The expected stop-sbottom contribution to the $T$ parameter 
 is constrained by naturalness to lie above the curve.}
\label{fig:T_stop}
\end{figure}

The \textit{stop-sbottom} contributions in the zero left-right mixing limit 
are given by
\begin{align}
  T_{\text{st-sb}} &= 6\,A(m_{\tilde{t}_{L}},m_{\tilde{b}_{L}})
    \approx \frac{m_{t}^{4}}{32\pi^{2}\alpha_{\rm em}v^{2}m_{\tilde{t}_{L}}^{2}}
    \approx 0.05 \left(\frac{500~\text{GeV}}{m_{\tilde{t}_{L}}}\right)^{2},
\\
  S_{\text{st-sb}} &= F(m_{\tilde{b}_{L}},m_{\tilde{b}_{L}})
      - F(m_{\tilde{t}_{L}},m_{\tilde{t}_{L}})
    \approx -\frac{1}{12\pi} \frac{m_{t}^{2}}{m_{\tilde{t}_{L}}^{2}}
    \approx -0.003 \left(\frac{500~\text{GeV}}{m_{\tilde{t}_{L}}}\right)^{2},
\end{align}
where the approximate expressions follow from $m_{\tilde{t}_{L}}^{2} 
- m_{\tilde{b}_{L}}^{2} = m_{t}^{2}$ in the large $m_{\tilde{t}_{L}}^{2}$ 
limit.  While the contribution to $S$ is always negligibly small, this is 
not the case, as is well known, for the contribution to $T$.  Anticipating 
a $\tan\beta$-dependent upper bound on the stop masses from naturalness 
considerations (see Section~\ref{sec:NC}), we show in Fig.~\ref{fig:T_stop} 
the minimum value of $T_{\text{st-sb}}$ when the stop masses are taken 
at this boundary.%
\footnote{Taking into account the left-right mixing may help reduce 
$T_{\text{st-sb}}$, although not dramatically, because the corresponding 
$A$-term parameter, $A_{t}$, will be subject to the same naturalness 
bound as $m_{\tilde{Q}}$ and $m_{\tilde{t}_{R}}$.}
This strongly reinforces the conclusion that $\tan{\beta}$ cannot be too 
large.  If the Higgs and stop-sbottom sectors were the only contributions 
to $S$ and $T$, any value of $\tan{\beta}$ above 5 or so would be almost 
excluded.

\begin{figure}[ptb]
\begin{center}
  \includegraphics[width=8.5cm]{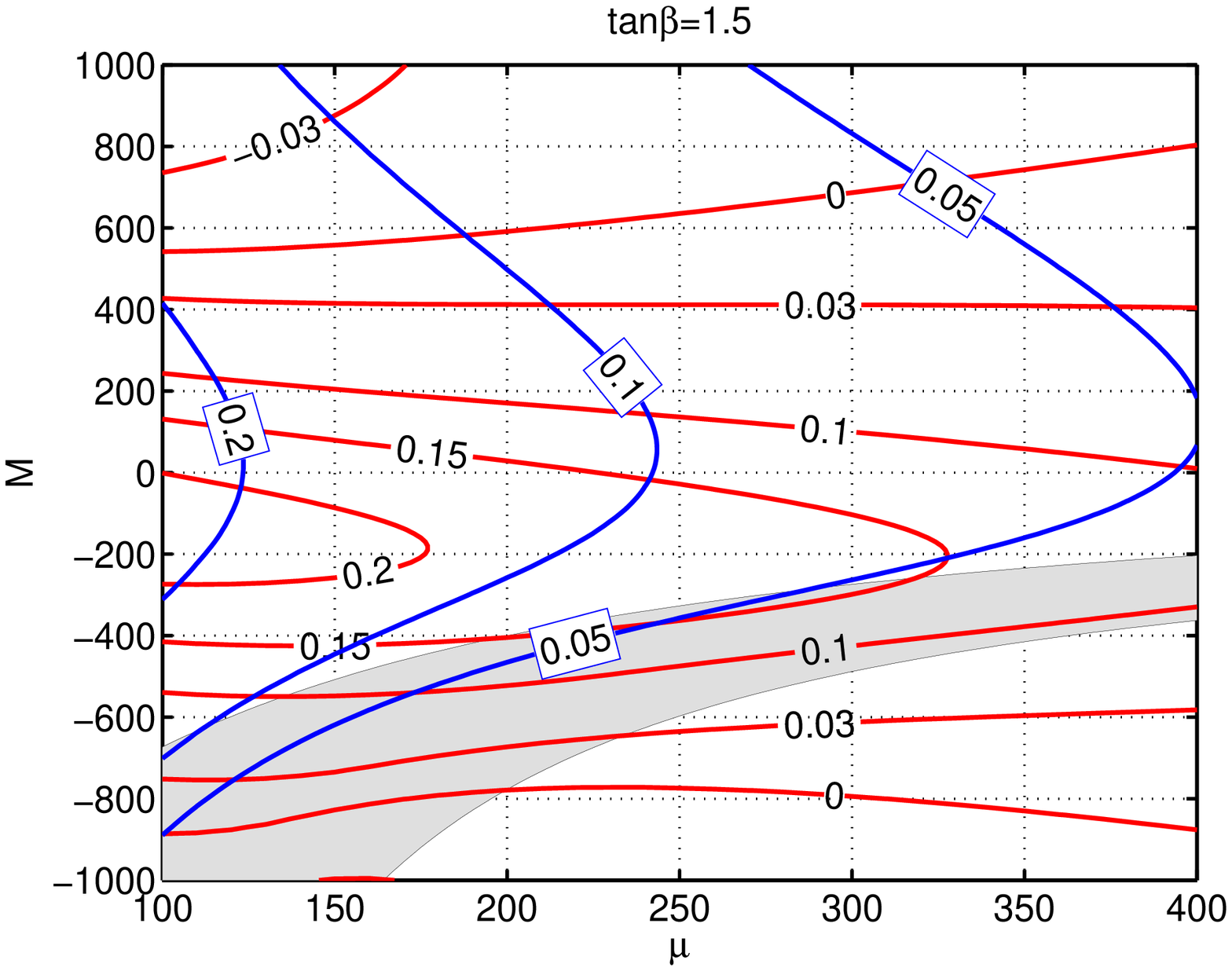} \includegraphics[width=8.5cm]{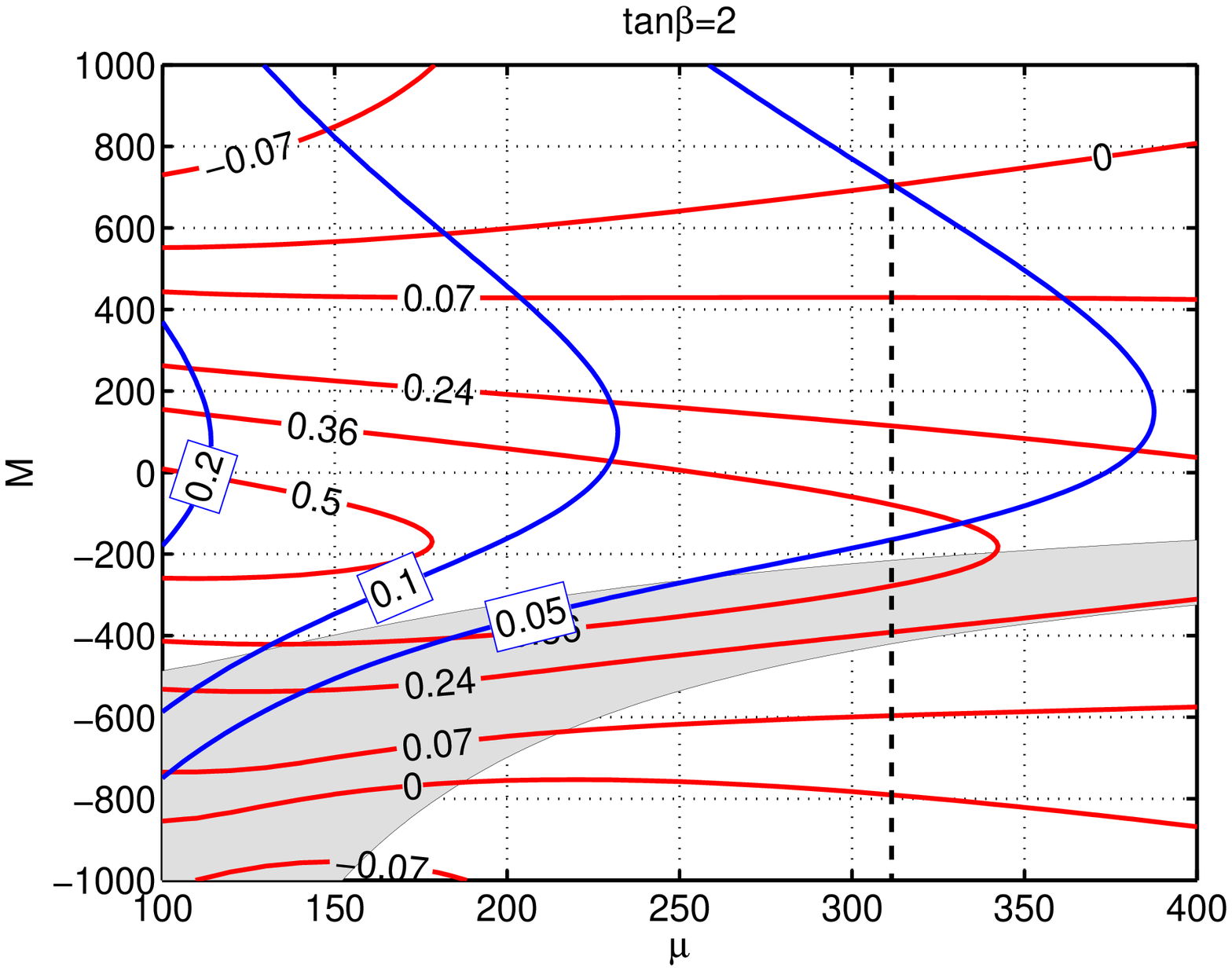}
  \includegraphics[width=8.5cm]{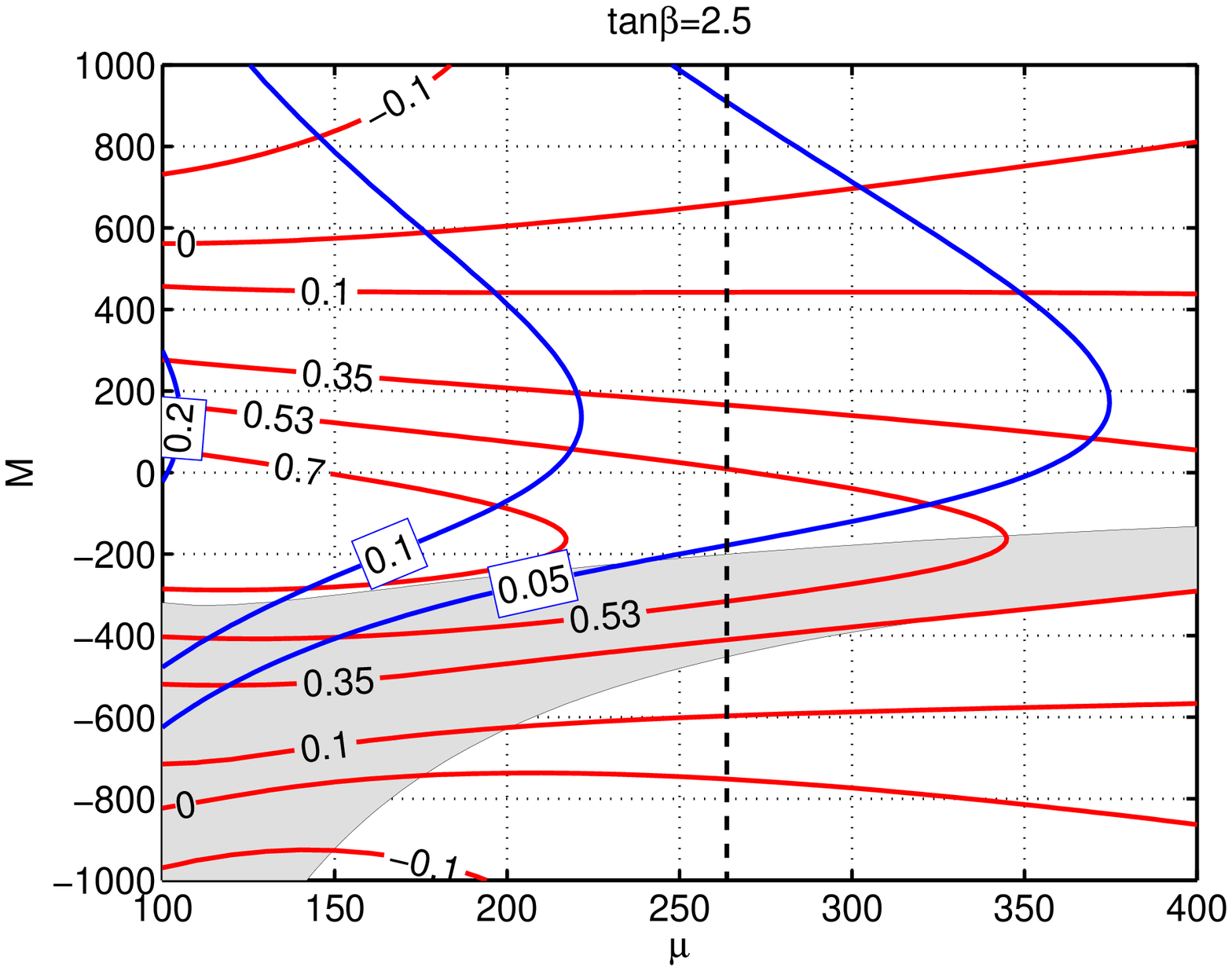} \includegraphics[width=8.5cm]{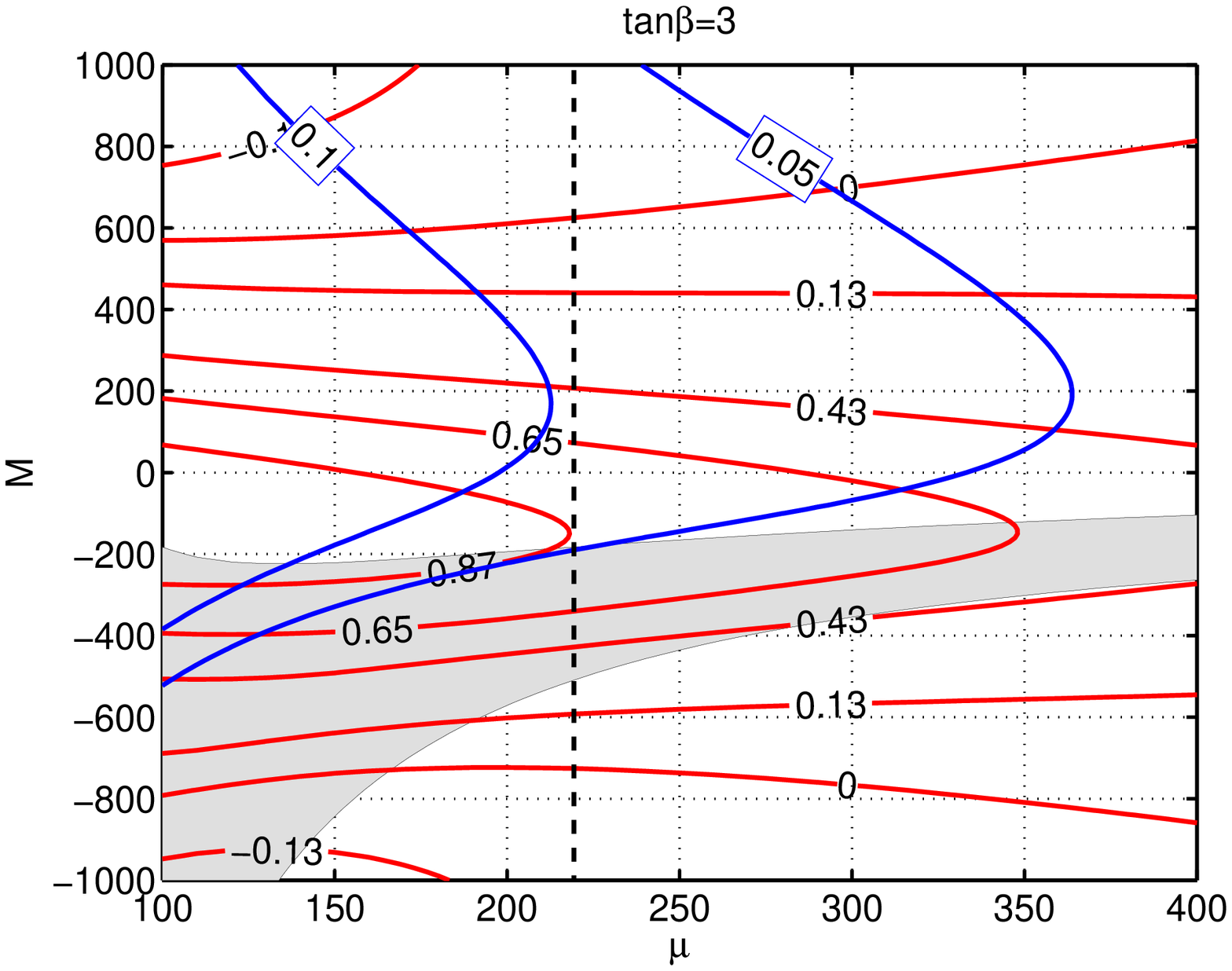}
\end{center}
\caption{The contributions from the Higgsinos to the $T$ parameter (red 
 color; contour values not framed) and the $S$ parameter (blue color; 
 contour values framed).  Vertical dashed lines represent an upper limit 
 on $\mu$ coming from naturalness.  Shaded regions indicate the regions in 
 which the lightest neutralino is lighter than $m_Z/2 \simeq 45~{\rm GeV}$.}
\label{fig:EWPT_Higgsinos}
\end{figure}

We finally consider the contributions from the Higgsino sector described 
in the previous Section.  The couplings to the gauge bosons are
\begin{align}
  \mathcal{L}_{\text{int}} =
    & \frac{g}{2} W_{\mu}^{+} \left( -\bar{\Psi}\gamma^{\mu}N_{1} 
      + \bar{\Psi}\gamma^{\mu}\gamma^{5}N_{2} \right) + \text{h.c.}
\nonumber\\
    & + \frac{g}{2}W_{\mu}^{3} \left( \bar{\Psi}\gamma^{\mu}\Psi
      + \bar{N}_{1}\gamma^{\mu}\gamma^{5}N_{2} \right) 
      + \frac{g^{\prime}}{2} B_{\mu} 
      \left( \bar{\Psi}\gamma^{\mu}\Psi 
      - \bar{N}_{1}\gamma^{\mu}\gamma^{5}N_{2} \right),
\end{align}
where $\Psi$ represents the chargino.  The contributions of Higgsinos to 
$T$ and $S$ can be written in terms of the mixing matrix $V$ as follows:
\begin{align}
  T_{\text{Higgsinos}} =
    & \sum\nolimits_{a=1}^{3} (V_{1a})^{2} \tilde{A}(\mu,m_{a})
      + (V_{2a})^{2} \tilde{A}(\mu,-m_{a})
\nonumber\\
    & - \frac{1}{2} \sum\nolimits_{a,b=1}^{3} (V_{1a}V_{2b}+V_{1b}V_{2a})^{2}
        \tilde{A}(m_{a},-m_{b}),
\\
  S_{\text{Higgsinos}} =
    & \frac{1}{2} \sum\nolimits_{a,b=1}^{3} (V_{1a}V_{2b}+V_{1b}V_{2a})^{2}
        \tilde{F}(m_{a},-m_{b}) - \tilde{F}(\mu,\mu),
\end{align}
(see Appendix~\ref{app:A} for the definitions of the functions $\tilde{A}$ and 
$\tilde{F}$).  These contributions are shown in Fig.~\ref{fig:EWPT_Higgsinos} 
in the $\mu$-$M$ plane. 
Also shown in the plot is the region (shaded) in which the lightest 
neutralino mass is less than half the $Z$ boson mass: $m_\chi < m_Z/2$. 
Even though its coupling to the $Z$ boson is suppressed, this mass 
range is inconsistent with the negative searches from the LEP.  In 
view of Figs.~\ref{fig:EWPT_scalars} and \ref{fig:T_stop}, as well 
as the mostly positive contributions to $S$ and $T$ in the plots of 
Fig.~\ref{fig:EWPT_Higgsinos}, we conclude that most of the $\mu$-$M$ plane 
is allowed for $\tan{\beta} \simlt 3$, except for a $\tan{\beta}$-dependent 
strip around $M=0$ where $T$ becomes too large.

The dependence of $T_{\text{Higgsinos}}$ on $\tan\beta$ 
can be approximated by%
\footnote{This relation can be understood easily in two limiting cases: 
1) for $\tan\beta \approx 1$, in which case we can compute $T$ perturbatively 
expanding in the $\mathcal{M}_{13}$ entry of the Higgsino mass matrix. 
2) in the decoupling limit of heavy Higgsinos, in which case integrating 
them out produces a dimension 6 operator $|H_{1}^{\dagger}D_{\mu}H_{1} 
- H_{2}^{\dagger}D_{\mu}H_{2}|^{2}$.  However, Eq.~(\ref{eq:Ta}) apparently 
holds with reasonable accuracy even in the general case of light Higgsinos 
and/or $\tan\beta - 1$ being not so small.}
\begin{equation}
  T_{\text{Higgsinos}} \approx {\cal F}(\tan\beta)\, \tilde{T}(\mu,M),
\qquad
  {\cal F}(t) = \left(\frac{t^{2}-1}{t^{2}+1}\right)^{2},
\label{eq:Ta}
\end{equation}
where $\tilde{T}(\mu,M)$ is some function of $\mu$ and $M$.  To make 
this scaling evident, the values for the contours of $T$ in each 
panel of Fig.~\ref{fig:EWPT_Higgsinos} are chosen such that they are 
proportional to the contours of the $\tan\beta = 1.5$ panel, scaled by 
a factor ${\cal F}(\tan\beta)/{\cal F}(1.5)$.  Even for $\tan\beta > 3$ 
there remain two strips around $M \approx 600~{\rm GeV}$ and $M \approx 
-800~{\rm GeV}$ where $T_{\text{Higgsinos}}$ is sufficiently small to be 
allowed (especially combined with the positive $S$ parameter).  However, 
taking into account the contribution $T_{\text{st-sb}}$, which starts 
being increasingly problematic for $\tan\beta \simgt (4\!\sim\!5)$, as 
well as the fact the allowed region for $\mu$ from naturalness rapidly 
shrinks for $\tan\beta \simgt (4\!\sim\!5)$, we conclude that the region
\begin{equation}
  \tan\beta \simlt 3,
\end{equation}
is preferred in the $\lambda$SUSY model.

\section{Dark Matter for Heavy Gauginos}
\label{sec:DM}

There are several possibilities in $\lambda$SUSY for the Lightest Supersymmetric 
Particle (LSP), corresponding to differing LHC signals and DM candidates. 
In this section we study the particular case that the LSP is dominantly 
Higgsino, having only small gaugino components.   In the previous section 
we have shown that decoupling the gauginos leads to a large region of parameter 
space consistent with EWPT, and in this section we find that the presence 
of the singlet Higgsino allows the interesting possibility of Higgsino DM, 
whereas in the MSSM it is excluded.

We now compute the thermal relic abundance of the lightest Higgsino $\chi$. 
We use the standard formalism from Ref.~\cite{Kolb:1988aj}.  The freeze-out 
point is given in terms of the scaled inverse temperature $x = m_\chi/T$:
\begin{equation}
  x_{f} 
  = \ln\frac{0.038\, g_{\chi}\, m_{\text{Pl}}\, m_{\chi} 
      \langle\sigma v_{\text{rel}}\rangle}{g_{\ast}^{1/2} x_{f}^{1/2}},
\end{equation}
where $m_{\text{Pl}} \simeq 1.22 \times 10^{19}~{\rm GeV}$, $g_{\chi} = 2$, 
$g_{\ast}$ is the number of SM degrees of freedom relativistic at the 
time of freeze-out ($g_{\ast} = 86.25$ for $m_{b} \ll T_{f} \simlt m_{W}$), 
and $\langle\sigma v_{\text{rel}}\rangle$ is the thermal averaged annihilation 
cross section ($\chi+\chi \rightarrow {\rm all}$), which enters the 
Boltzmann equation for the $\chi$ number density $n$:
\begin{equation}
  \frac{dn}{dt} + 3Hn 
  = -\langle\sigma v_{\text{rel}}\rangle (n^{2}-n_{\text{eq}}^{2}).
\end{equation}
At freeze-out, $\chi$ is nonrelativistic and the cross section can be 
expanded in $v$:
\begin{equation}
  \sigma v_{\text{rel}} = a + b v_{\text{rel}}^{2},
\qquad
  \langle\sigma v_{\text{rel}}\rangle = a + 6 b /x.
\end{equation}
Defining the annihilation integral
\begin{equation}
  J(x_{f}) = \int_{x_{f}}^{\infty} 
    \frac{\langle\sigma v_{\text{rel}}\rangle}{x^{2}} dx
  \approx \frac{a}{x_{f}} + \frac{3b}{x_{f}^{2}},
\end{equation}
the present-day mass density of $\chi$ is given by
\begin{equation}
  \Omega_{\chi} h^{2} 
    \simeq \frac{1.07 \times 10^{9}~\text{GeV}^{-1}}
    {g_{\ast}^{1/2} m_{\text{Pl}}\, J(x_{f})}.
\end{equation}

There are several caveats in the analysis described above. 1) We ignore 
possible coannihilations.  This will be valid if all the other neutralinos 
and charginos have masses at least a few$\times T_{f}$ higher than $m_\chi$. 
This treatment is indeed justified in the parameter region of interest to us, 
but may not be in general.  2) We do not give a careful treatment of 
near-threshold situations.  Indeed, there are important thresholds associated 
with the $W$ and $Z$ bosons.  Near these thresholds, our calculation may 
be subject to larger errors, and a more careful treatment should proceed 
along the lines of Ref.~\cite{Griest:1990kh}.

Let us now calculate various annihilation processes in turn.  We first 
calculate $\chi+\chi \rightarrow f\bar{f}$ via $s$-channel $Z$ exchange. 
The $\chi$ coupling to $Z$ is given by
\begin{equation}
  \frac{g}{2c_{\text{w}}}\frac{\kappa_{\chi}}{2}
    Z_{\mu}\bar{\chi}\gamma^{\mu}\gamma^{5}\chi,
\qquad
  \kappa_{\chi} = 2V_{1\chi}V_{2\chi}.
\end{equation}
The resulting annihilation cross section in the approximation of massless final 
states is $\sigma_{f\bar{f}}v_{\text{rel}} = b_{f\bar{f}}v_{\text{rel}}^{2}$:
\begin{equation}
  b_{f\bar{f}} = 7.31 \kappa_{\chi}^{2} \frac{g^{4}}{32\pi c_{\text{w}}^{4}}
    \frac{m_{\chi}^{2}}{(4m_{\chi}^{2}-m_{Z}^{2})^{2}},
\end{equation}
where $7.31$ is the sum of $g_{V}^{2}+g_{A}^{2}$ over all SM quarks and 
leptons except for the top quark.  The coupling $\kappa_{\chi}$ is generically 
suppressed, especially in the lower right corner of the $\mu$-$M$ plane. 
For this reason, large values of $\Omega_{\chi} h^{2}$ are attained for 
modest values of $m_{\chi}$.

Above the $W$ thresholds, we have to take into account the annihilation 
into $W^{+}W^{-}$ proceeding via $s$-channel $Z$ exchange and $t$,$u$-channel 
chargino exchange.  The amplitudes for these processes are given by
\begin{align}
  & \frac{g^{2}}{2} \kappa_{\chi} 
      \left( \bar{v}_{2}\gamma_{\lambda}\gamma_{5}u_{1} \right) 
      \frac{i}{s-m_{Z}^{2}}
\nonumber\\
  & \times\left[ \eta^{\mu\nu}(k_{-}-k_{+})^{\lambda}
      - \eta^{\lambda\nu}(k_{+}+2k_{-})^{\mu}
      + \eta^{\lambda\mu}(2k_{+}+k_{-})^{\nu} \right]
      \varepsilon_{\mu}^{\ast}(k_{+}) \varepsilon_{\nu}^{\ast}(k_{-}),
\end{align}
and 
\begin{equation}
  \frac{g^{2}}{4} \left[ \bar{v}_{2}\gamma^{\nu}P
    \frac{i(\slashed{k}_{+}-\slashed{p}_{1}+\mu)}{(k_{+}-p_{1})^{2}-\mu^{2}}
    \gamma^{\mu}Pu_{1} - (1 \leftrightarrow 2) \right] 
    \varepsilon_{\mu}^{\ast}(k_{+}) \varepsilon_{\nu}^{\ast}(k_{-}),
\qquad 
  P = V_{1\chi}-V_{2\chi}\gamma^{5},
\end{equation}
respectively, where $u_{1,2}$ are the initial state spinors with momenta 
$p_{1,2}$, $\bar{v}_{i} \equiv u_{i}C$ with $C$ being the charge conjugation 
matrix.  Since the $Z$-exchange amplitude vanishes at rest, only the latter 
amplitude contributes to the $a$ coefficient of the resulting cross section 
$\sigma_{WW}v_{\text{rel}} = a_{WW} + b_{WW}v_{\text{rel}}^{2}$:
\begin{equation}
  a_{WW} = \frac{g^{4}}{32\pi} 
    \left(1-\frac{m_{W}^{2}}{m_{\chi}^{2}}\right)^{3/2}
    \frac{(V_{1\chi}^{2}+V_{2\chi}^{2})^{2} m_{\chi}^{2}}
      {(m_{\chi}^{2}+\mu^{2}-m_{W}^{2})^{2}}.
\end{equation}
However, $b_{WW}$ needs to be included since it turns out that in this case 
typically $b\gg a$ and the correction to the cross section at freeze-out 
is of order 1.  Both the $Z$-exchange and chargino-exchange amplitudes 
contribute to $b_{WW}$.  The resulting expression is very long, and we 
do not present it here.

Finally, above the $Z$ thresholds, we have to take into account the 
annihilation into $ZZ$ proceeding via $t$,$u$-channel exchanges of 
all the three neutralinos $\chi_{a}$ ($\chi \equiv \chi_{1}$) with the 
amplitude:
\begin{equation}
  \frac{g^{2}}{4c_{\text{w}}^{2}} \sum_{a=1}^{3}\kappa_{a}^{2}
    \left[ \bar{v}_{2}\gamma^{\nu}\gamma^{5}
    \frac{i(\slashed{k}_{1}-\slashed{p}_{1}+m_{a})}
      {(k_{1}-p_{1})^{2}-m_{a}^{2}} \gamma^{\mu} \gamma^{5} u_{1}
    - (1 \leftrightarrow 2) \right] 
    \varepsilon_{\mu}^{\ast}(k_{1}) \varepsilon_{\nu}^{\ast}(k_{2}),
\qquad
  \kappa_{a} = V_{1\chi}V_{2a} + V_{1a}V_{2\chi}.
\end{equation}
The \thinspace$a$ coefficient of the resulting cross section $\sigma_{ZZ}$ 
can be given explicitly:
\begin{equation}
  a_{ZZ} = \frac{g^{4}}{64\pi c_{\text{w}}^{4}} 
    \left(1-\frac{m_{Z}^{2}}{m_{\chi}^{2}}\right)^{3/2} \left[ \sum_{a=1}^{3} 
    \frac{\kappa_{a}^2\, m_{\chi}}{m_{\chi}^{2}+m_{a}^{2}-m_{Z}^{2}} \right]^2.
\end{equation}
However, $b_{ZZ}$ is again needed, which we do not present here.

In our computation, we do not include the $t$-channel sfermion exchange 
contributions to the $f\bar{f}$ final states or the $s$-channel Higgs-boson 
exchange contributions to the $WW$ and $ZZ$ final states, which are generically 
suppressed and depend on free parameters other than $\mu$, $M$, $\lambda$ 
and $\tan\beta$.  We also do not consider final states involving the Higgs 
boson(s), whose thresholds are higher than the ones discussed above.

\begin{figure}[ptb]
\begin{center}
  \includegraphics[width=8.5cm]{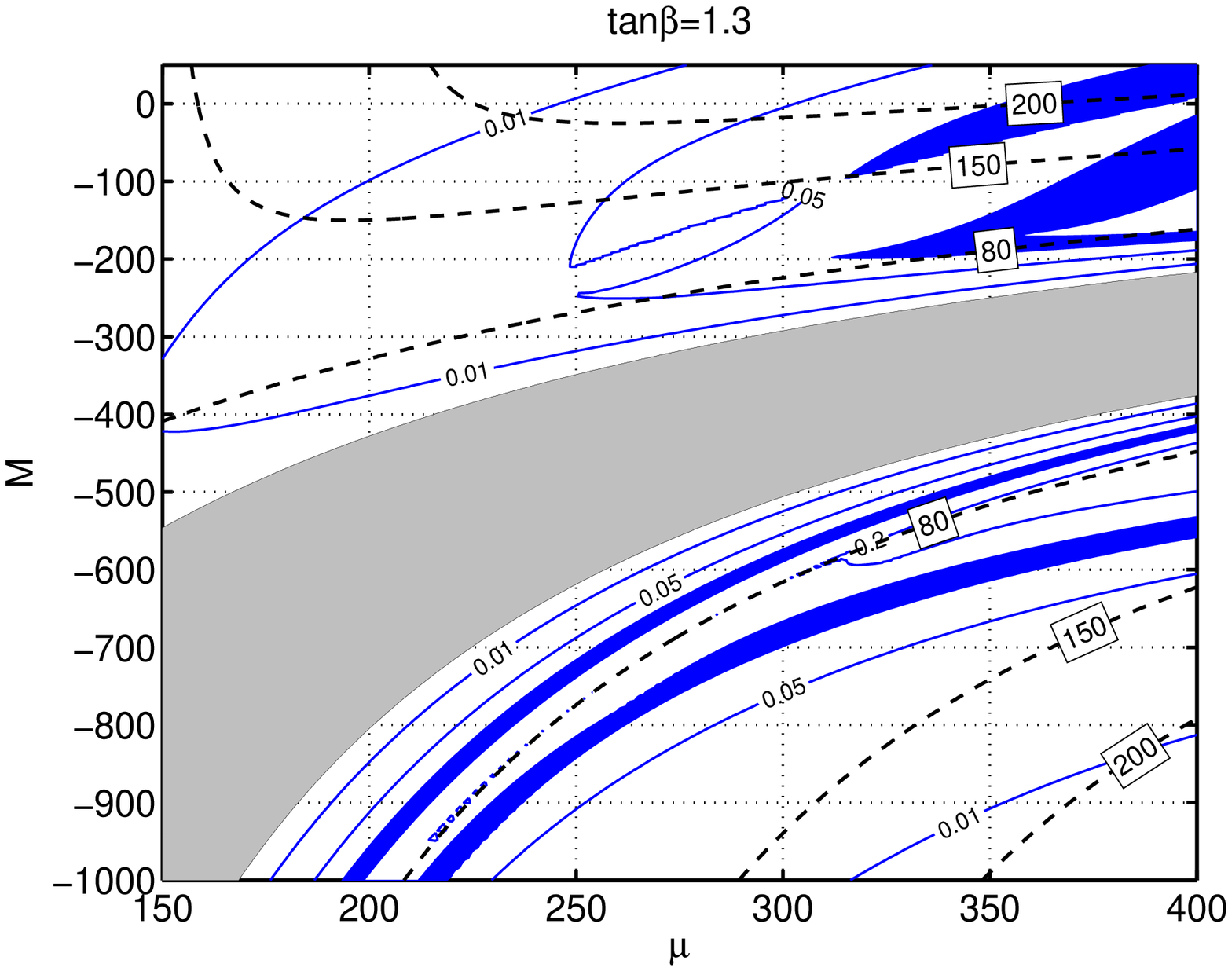}
  \includegraphics[width=8.5cm]{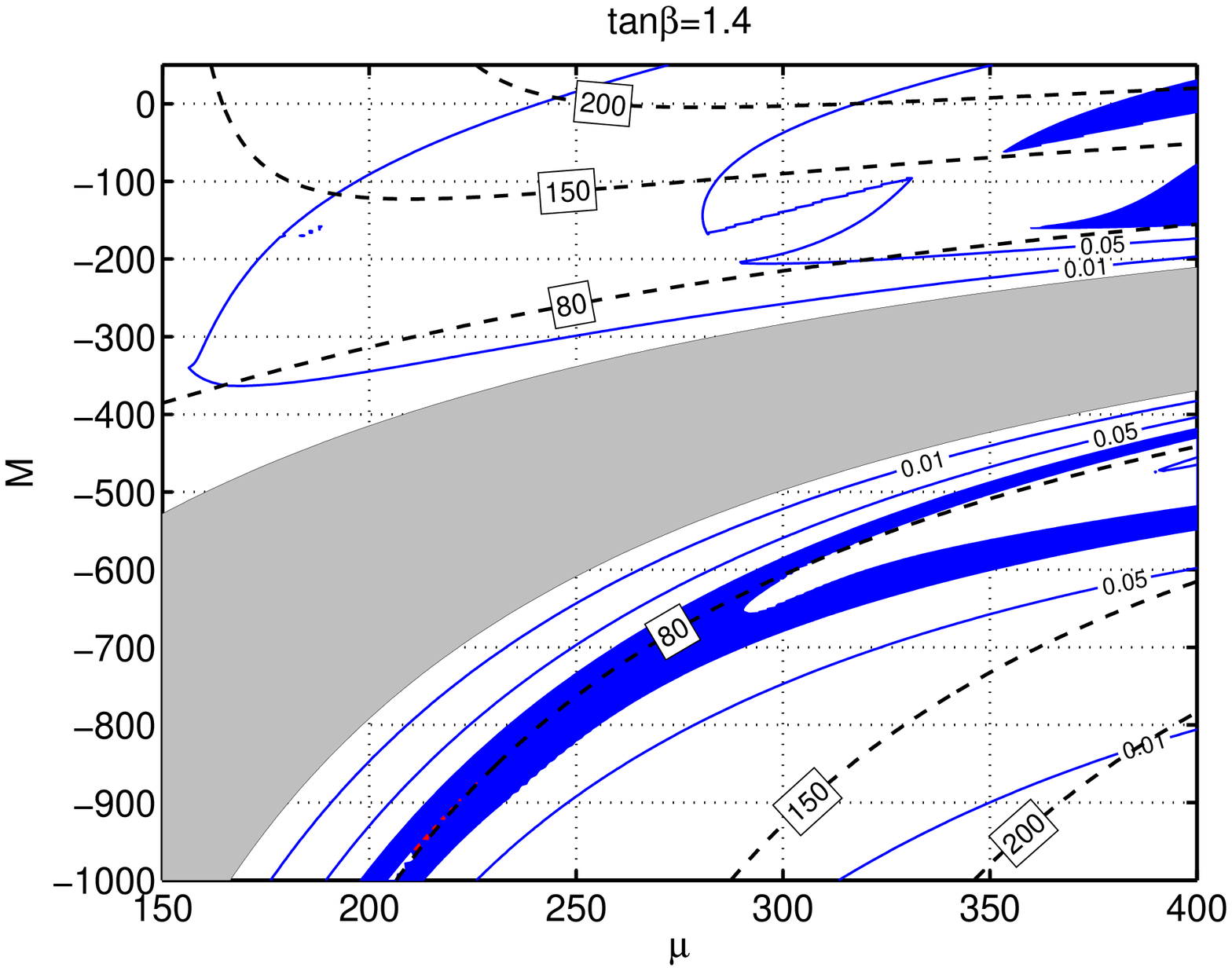}
  \includegraphics[width=8.5cm]{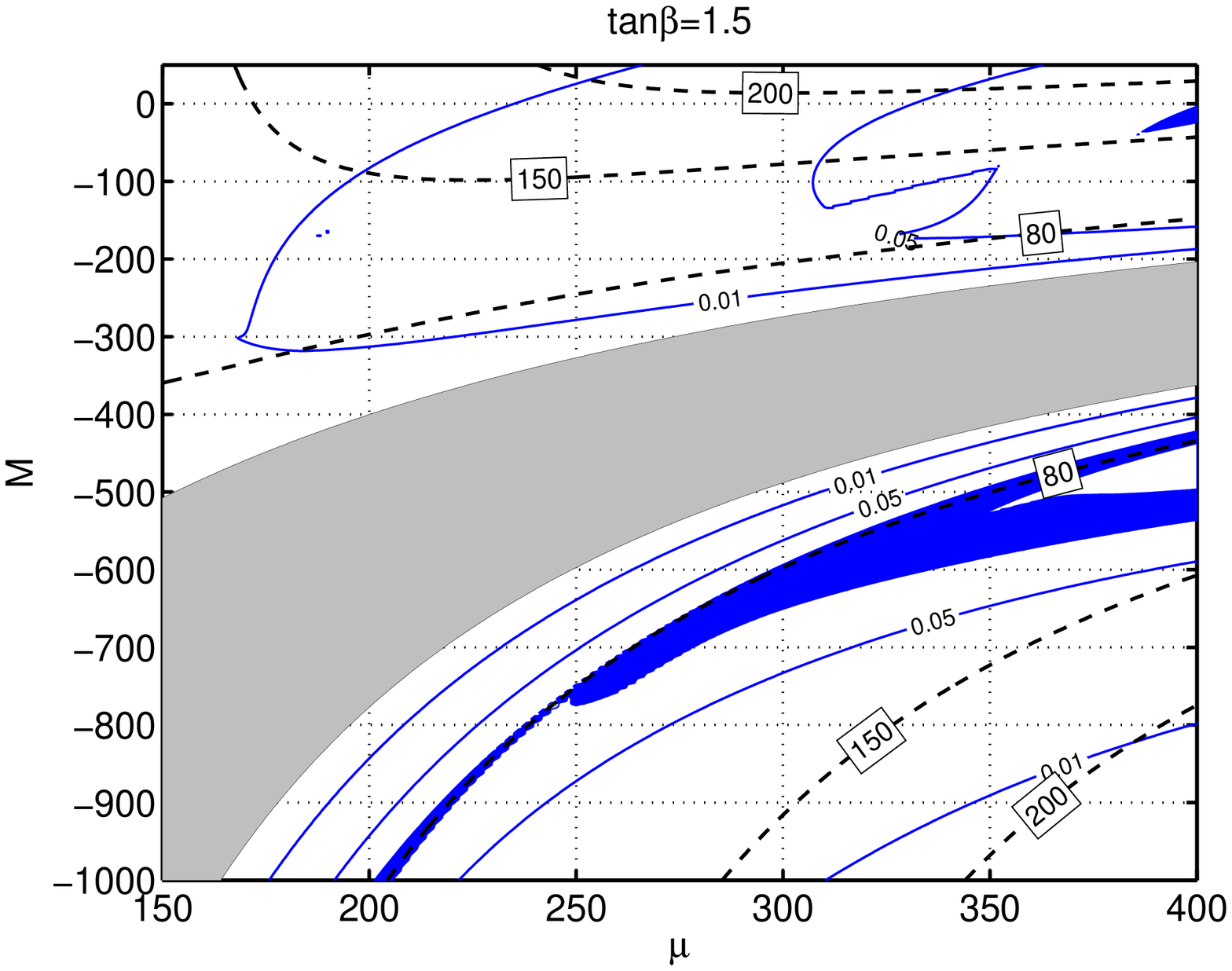}
  \includegraphics[width=8.5cm]{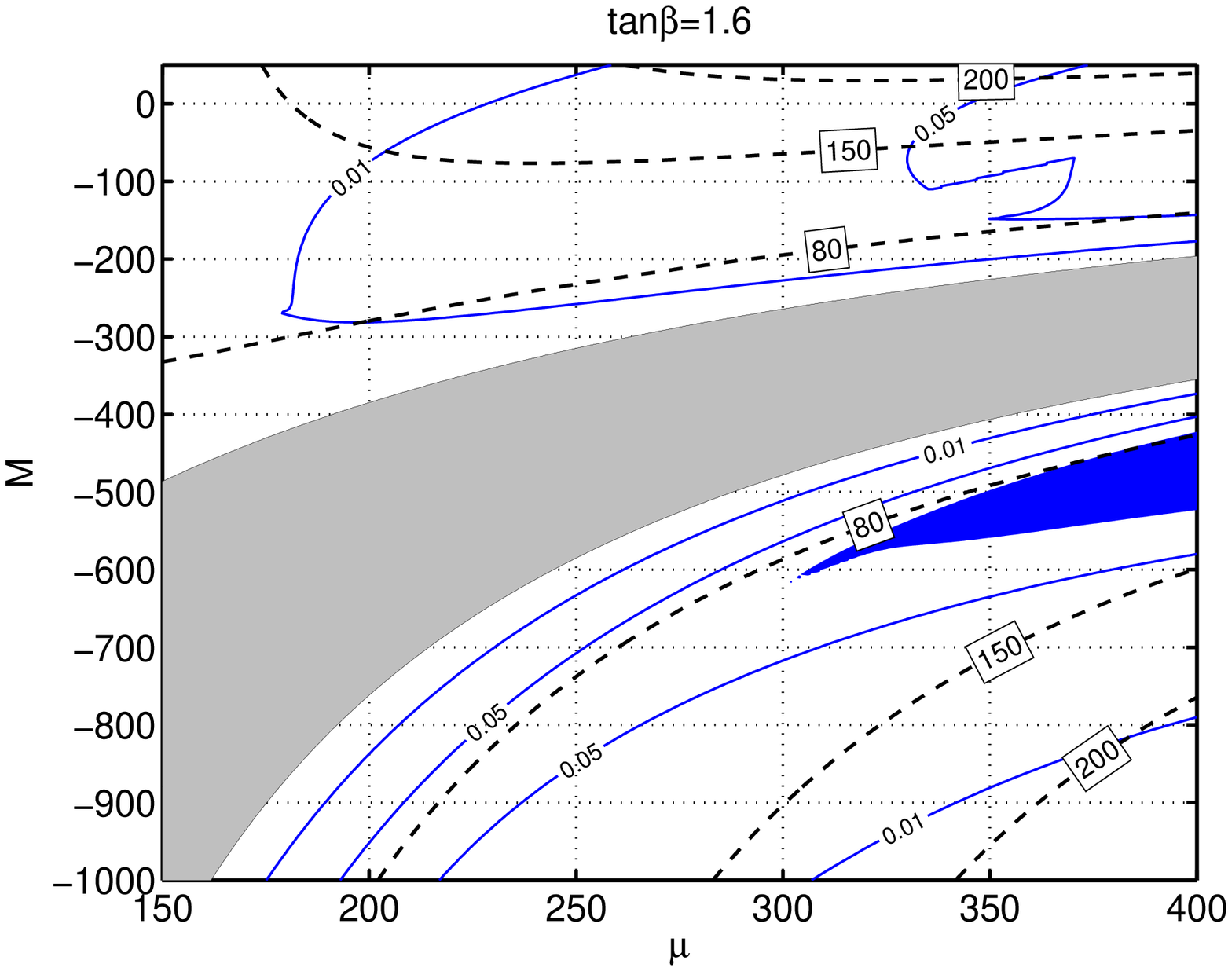}
\end{center}
\caption{The lightest neutralino relic density $\Omega_{\chi} h^{2}$ 
 for four values of $\tan\beta$.  The contours for $\Omega_{\chi} h^{2} 
 = 0.01, 0.05$ and $0.2$ are denoted by the blue (thin solid) lines, 
 while the blue (darkest) shading indicates regions with $0.09 \simlt 
 \Omega_{\chi} h^{2} \simlt 0.13$, corresponding to the $95\%$CL region 
 from WMAP~\cite{Spergel:2003cb}.  Gray shading indicates regions where 
 $m_{\chi} < m_{Z}/2$.  The contours for the mass of the LSP, $m_\chi 
 = 80, 150$ and $200~{\rm GeV}$, are also superimposed by dashed lines.}
\label{fig:DM_plot}
\end{figure}

We now report the results of our numerical computations for the DM abundance. 
In Fig.~\ref{fig:DM_plot} we show plots for the DM relic abundance in the 
$\mu$-$M$ plane for 4 different values of $\tan\beta \equiv t = 1.3, 1.4, 1.5$ 
and $1.6$.
In most of the parameter space for $t \approx (1\!\sim\!3)$, which we are 
interested in, the thermal relic abundance is much smaller than the observed 
DM abundance ($< 10\%$).  Around $t = 1.4$, there are regions where the 
abundance has the observed value.  The plots show the situation for 4 
representative values of $t$ around $1.4$.  Going to smaller $t$ is disfavored 
by the EWPT, while for larger $t$ the DM region gets pushed towards larger 
values of $\mu$, which are disfavored by naturalness.  We conclude that 
the new possibility for DM discussed here is compatible with the EWPT and 
naturalness for $\tan\beta \simlt 1.7$.

In the MSSM with decoupled gauginos, the lightest neutralino is a 
pseudo-Dirac Higgsino of mass $\mu$.  Such a Higgsino has a large coupling 
to the $Z$ and, since naturalness places a bound on how large $\mu$ can be, 
the Higgsinos annihilate easily in the early universe contributing little 
to the DM.  How does the LSP Higgsino of $\lambda$SUSY avoid this?  There 
are two possibilities.  In the DM regions with low $|M|$, visible for 
the low values of $\tan\beta$ in Fig.~\ref{fig:DM_plot}, $\chi$ is mainly 
singlet and this reduces the $Z$ coupling.  On the other hand, in the high 
$|M|$ DM regions, the $H_1$ and $H_2$ components of the Majorana $\chi$ 
lead to a partial cancellation of the $Z$ coupling, again allowing 
substantial DM.

We finally consider the direct detection cross section of $\chi$.  The $\chi$ 
coupling to the lightest Higgs is given by
\begin{equation}
  \frac{\lambda\delta}{\sqrt{2}} \bar{\chi}\chi h,
\qquad
  \delta=V_{3\chi}(V_{1\chi}\cos(\alpha-\pi/4)+V_{2\chi}\cos(\alpha+\pi/4)),
\end{equation}
while the coupling of $h$ to the quarks is given by
\begin{equation}
  \frac{1}{\sqrt{2}v} h 
    \left( \frac{\cos\alpha}{\sin\beta} m_{u}\bar{q}_{u}q_{u}
      - \frac{\sin\alpha}{\cos\beta} m_{d}\bar{q}_{d}q_{d} \right).
\end{equation}
The nucleon cross section is then given by
\begin{equation}
  \sigma_{h}(\chi\mathcal{N}\rightarrow\chi\text{\thinspace}\mathcal{N})
  = \frac{m_{r}^{2}}{\pi} 
    \left(\frac{\lambda\delta}{m_{h}^{2}v}\right)^{2} 
    \left( X_u \frac{\cos\alpha}{\sin\beta} 
      - X_d \frac{\sin\alpha}{\cos\beta} \right)^{2} 
    m_{\mathcal{N}}^{2},
\label{eq:higgsexch}
\end{equation}
where $m_r \equiv m_\chi m_{\mathcal{N}}/(m_\chi + m_{\mathcal{N}})$ is 
the reduced mass, and $X_u$ and $X_d$ are certain linear combinations of 
nucleon matrix elements, which we conservatively take as $X_u \simeq 0.14$ 
and $X_d \simeq 0.24$~\cite{Drees:1993bu}.  The cross section on a nucleus, 
normalized to a nucleon, is (for $\lambda=2$):
\begin{equation}
  \sigma_{h}(\chi p\rightarrow\chi p)
  \equiv 10^{-44}~\text{cm}^{2}\, G,
\qquad
  G \simeq 200 \times \delta^{2} 
    \left( X_u \frac{\cos\alpha}{\sin\beta} 
      - X_d \frac{\sin\alpha}{\cos\beta} \right)^{2} 
    \left( \frac{300~\text{GeV}}{m_{h}} \right)^{4}.
\label{eq:higgsexchnum}
\end{equation}
Recall that for a DM mass in the $50-200~{\rm GeV}$ range of interest to 
us, the limit on the spin-independent direct detection cross section is 
$(2\!\sim\!3) \times 10^{-43}~{\rm cm}^{2}$~\cite{Akerib:2005kh}.  We have 
studied numerical values of the factor $G$ in the region of $\tan\beta 
= 1.3$~--~$3$ and found that it never exceeds $\simeq (20\!\sim\!30)$. 
Thus the cross section is not in contradiction with the existing limits, 
even without taking into account the fact that in most of the parameter 
space the LSP relic abundance is much below the observed DM density. 
We have further studied the factor $G$ for the regions of parameter 
space with $\tan\beta = 1.3$~--~$1.6$ where we get the observed relic 
density, i.e. the blue (darkest) regions in Fig.~\ref{fig:DM_plot}. 
\begin{figure}[ptb]
\begin{center}
  \includegraphics[width=8.5cm]{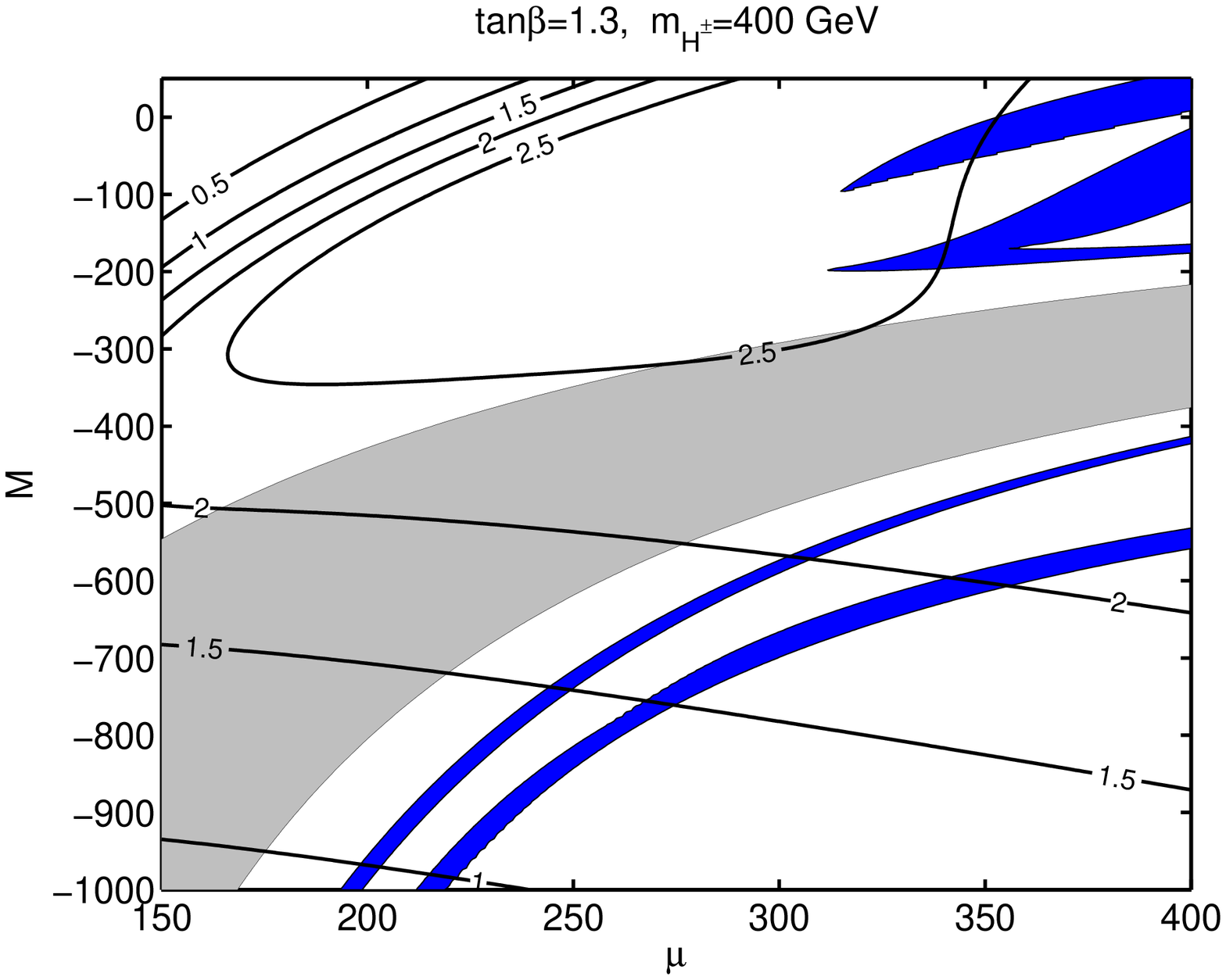}
  \includegraphics[width=8.5cm]{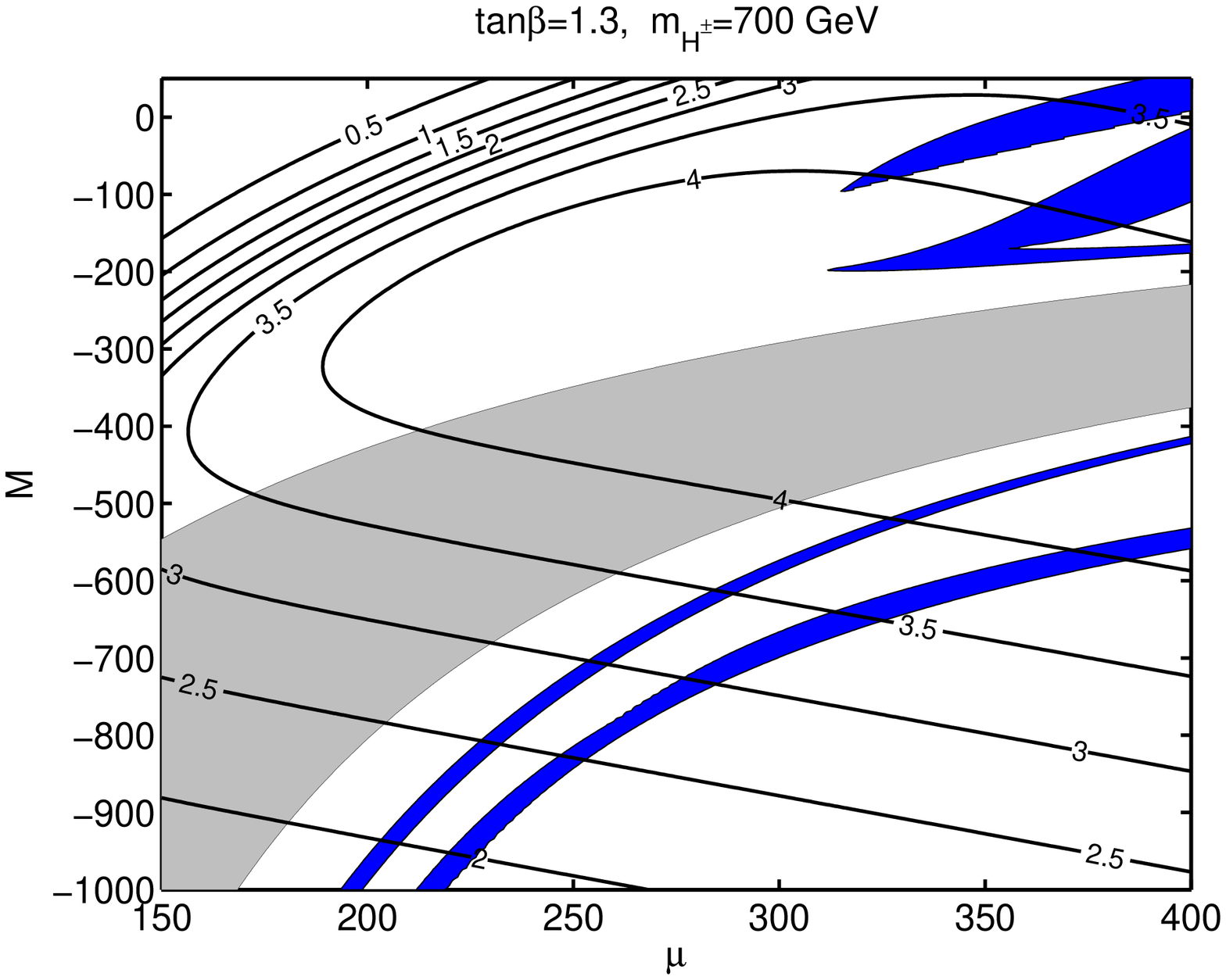}
  \includegraphics[width=8.5cm]{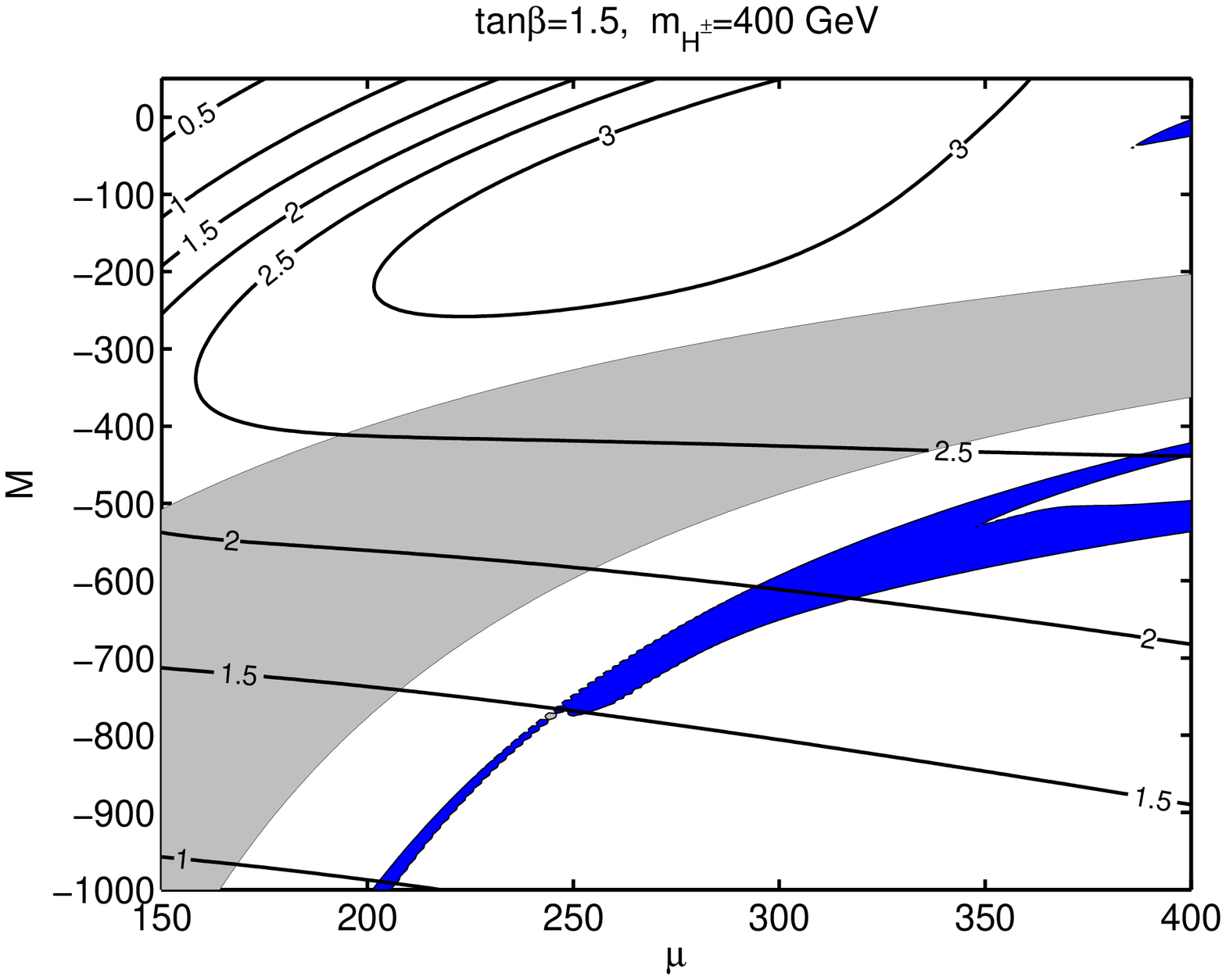}
  \includegraphics[width=8.5cm]{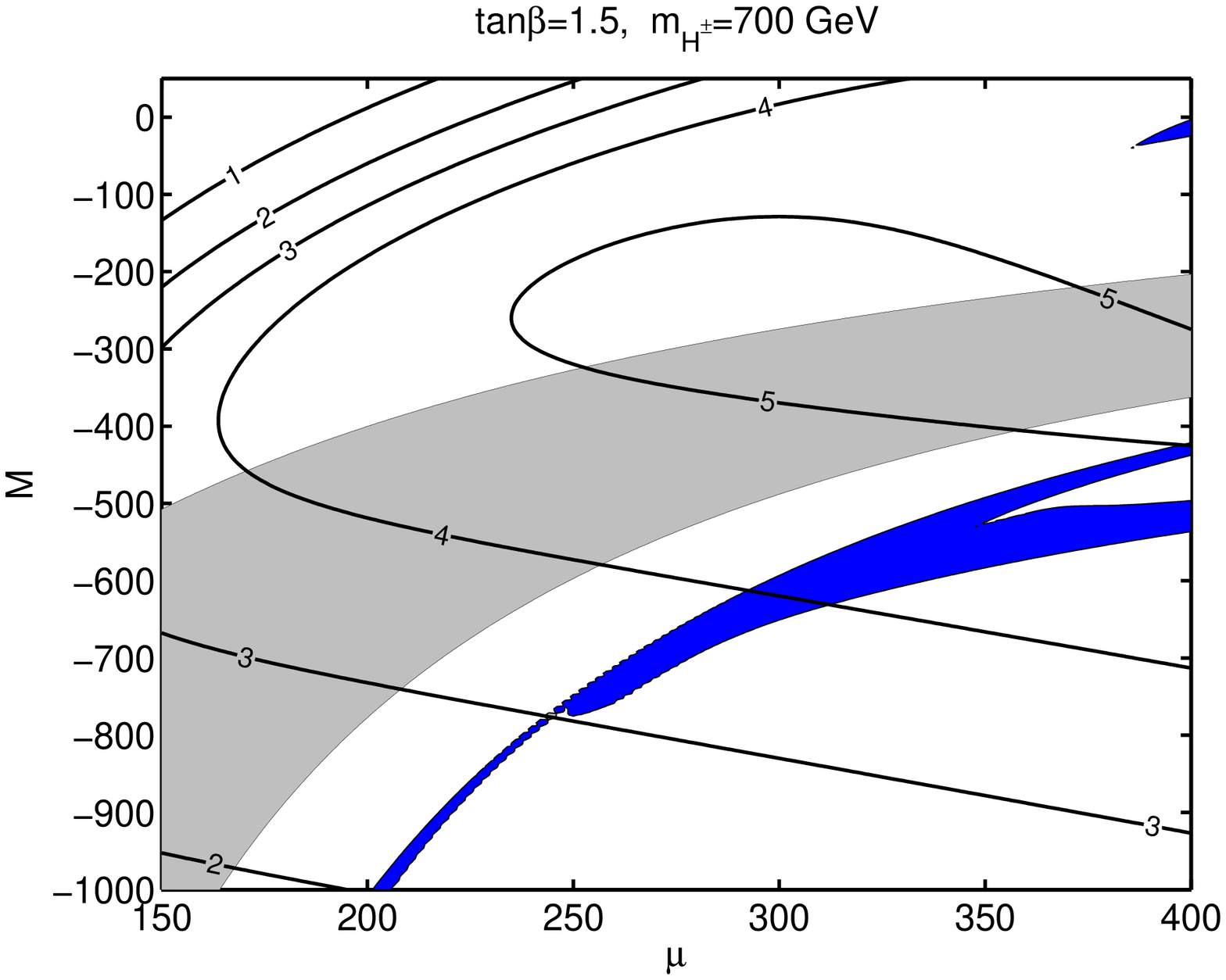}
\end{center}
\caption{Dark matter detection cross section $\sigma_{h}$ in units of 
 $10^{-44}~{\rm cm}^2$ (i.e. the factor $G$ in the text) in the $\mu$-$M$ 
 plane for $\tan\beta = 1.3, 1.5$ and $m_{H^\pm} = 400, 700~{\rm GeV}$. 
 The blue (darkest) regions indicate the regions for $0.09 \simlt 
 \Omega_{\chi} h^{2} \simlt 0.13$ (see Fig.~\ref{fig:DM_plot}), while 
 the gray shaded regions indicate the ones where $m_{\chi} < m_{Z}/2$.}
\label{fig:DM_sigma}
\end{figure}
We find that $G$ ranges from about $1$ to $5$, as depicted in 
Fig.~\ref{fig:DM_sigma} for the case of $m_{H^\pm} = 400, 700~{\rm 
GeV}$ for $\tan\beta = 1.3, 1.5$.  This puts the cross section 
(\ref{eq:higgsexchnum}) well within reach of experiments currently 
under way.

\section{Naturalness Constraints on Superparticle Masses}
\label{sec:NC}

As mentioned in the Introduction, making the lightest Higgs boson heavier 
helps in relaxing the naturalness constraints on the model, while allowing 
heavier superpartners.  It is of interest therefore to establish on 
quantitative grounds the upper bounds that can be set on physical 
superparticle masses.  This is of direct relevance also to understand the 
precise significance of the results obtained in the previous two sections. 

For this purpose, we consider an explicit simplified version of the 
$\lambda$SUSY model, defined by the following superpotential and the 
soft supersymmetry-breaking Lagrangian
\begin{align}
  W &= \lambda S H_{1} H_{2} + m H_{1} H_{2} + \frac{1}{2} M S^{2},
\\
  {\cal L}_{\rm soft} &= -m_{S}^{2}|S|^{2} 
    - \sum_{i=1,2} \mathbf{m}_{i}^{2} |H_{i}|^{2} 
    + (B m H_{1}H_{2} + {\rm h.c.}),
\end{align}
and we will study especially the case in which the soft breaking 
mass $m_{S}$ is taken large, consistently with naturalness.  (See 
Ref.~\cite{Nomura:2005rk} for an earlier analysis.)  We do not expect 
the bounds obtained in the generic $\lambda$SUSY model to be significantly 
different, or in any case more restrictive, than the ones valid in this 
specific case.

The scalar potential takes the form of Eq.~(\ref{eq:pot}) with
\begin{equation}
  \mu_{1}^{2} = \mathbf{m}_{1}^{2} + \mu^{2},
\qquad
  \mu_{2}^{2} = \mathbf{m}_{2}^{2} + \mu^{2},
\qquad
  \mu_{3}^{2} = Bm - \lambda M s,
\label{eq:Ms}
\end{equation}
\begin{equation}
  V(S) = (M^{2}+m_{S}^{2})|S|^{2} \equiv \mu_{S}^{2}|S|^{2},
\end{equation}
and the chargino mass parameter defined in Eq.~(\ref{eq:superpot}) 
is $\mu = m + \lambda s$, where $s$ is the background expectation value 
of $S$.  In this model, the stability condition (\ref{eq:stab}) implies 
also the conservation of $CP$ at the vacuum.

\subsection{Minimizing the potential}

\begin{figure}[ptb]
\begin{center}
  \includegraphics[width=8cm]{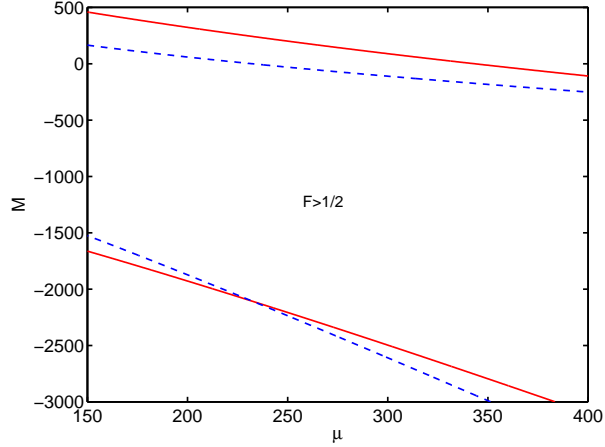}
\end{center}
\caption{The region with $F > 1/2$ in the $\mu$-$M$ plane, for $t=1.5$ 
 (between red solid lines) and for $t=3$ (between blue dashed lines).}
\label{fig:F}
\end{figure}

To understand the dependence of the minimum of the potential on various 
parameters, it is best to minimize first with respect to $v_{1}$ and $v_{2}$ 
for fixed $s$.  We then obtain
\begin{equation}
  V = \mu_{S}^{2} s^{2} - \frac{1}{\lambda^{2}} 
        \left( \mu_{3}^{2}(s) - \mu_{1}(s) \mu_{2}(s) \right)^{2}.
\end{equation}
This potential can in turn be minimized with respect to $s$, giving 
\begin{align}
  \lambda s 
  &= -\gamma (\mu_{3}^{2}-\mu_{1}\mu_{2})
   = -\gamma \frac{\lambda^{2}v^{2}}{t+t^{-1}},
\\
  \gamma 
  &= \frac{M+m(t+t^{-1})}{\mu_{S}^{2}+\lambda^{2}v^{2}},
\end{align}
where $t \equiv \tan{\beta}$.  Plugging this expression for $s$ into 
Eq.~(\ref{eq:Ms}) and then into (\ref{eq:lambdav}), we can obtain the 
expression for $v$ in terms of the original parameters of the model. 
Assuming that $s$ is small relative to $v$, which is the case for 
sufficiently large $\mu_{S}$, we find
\begin{equation}
  \lambda^{2}v^{2} \simeq 
    Bm \left(t+t^{-1}\right) - \left(\mathbf{m}_{1}^{2} + m^2\right) 
    - \left(\mathbf{m}_{2}^{2} + m^2\right) 
    + \lambda^{2}v^{2} \left( M + \frac{4m}{t+t^{-1}} \right) \gamma,
\end{equation}
which can be rewritten in the form
\begin{align}
  \lambda^{2}v^{2} &\simeq 
    F^{-1} \left[ Bm \left(t+t^{-1}\right) - \left(\mathbf{m}_{1}^{2} + m^2\right) 
      - \left(\mathbf{m}_{2}^{2} + m^2\right) \right],
\label{eq:lambda2-v2}
\\
  F &= 1 - \left( M + \frac{4m}{t+t^{-1}} \right) \gamma.
\end{align}
This gives, together with $t = \mu_{1}/\mu_{2}$, the explicit dependence 
of $v$ on the original parameters of the model, which in turn allows us 
to determine the naturalness constraints.  Note that the expression in 
the square bracket of Eq.~(\ref{eq:lambda2-v2}) is the one which we would 
obtain in the pure 2HDM without the singlet field $S$.  By taking $m_{S}$ 
sufficiently large, consistently with naturalness (see below), and 
restricting the range of other parameters, we shall always require 
that the factor $F$ be sufficiently close to unity so that we can 
neglect its presence.  This is consistent with our treatment of the 
EWPT in Section~\ref{sec:EWPT}, where the mixings between the singlet 
and doublet scalars were neglected.  $F > 1/2$ is the numerical 
condition that we shall take to constrain the general space of 
the parameters.  The regions $F > 1/2$ are depicted in Fig.~\ref{fig:F} 
in the $\mu$-$M$ plane for $t=1.5$ and $3$.  Here, the values of 
$m_{H^\pm}$ and $m_{S}^{2}$ are taken to saturate the naturalness 
upper bounds discussed in the next subsection (see Eqs.~(\ref{eq:F1-F2},%
~\ref{eq:mH-bound},~\ref{eq:mS-bound})), although using smaller values 
of $m_{H^\pm}$ leads to only negligible variations of the regions.

\subsection{Naturalness bounds}

We are interested in the naturalness constraints on $\mu_{3}$, which sets 
the bound on the Higgs sector through (\ref{eq:mA}) and on the chargino 
mass through (\ref{eq:mu-bound}).  We also want to know the bound on the 
stop masses and on the mass of the scalar $S$, which both affect, via one 
loop corrections, the masses $\mathbf{m}_{1}^{2}$ and $\mathbf{m}_{2}^{2}$, 
and therefore the masses $\mu_{1}^{2}$ and $\mu_{2}^{2}$.  As customary, 
we shall require that the logarithmic derivatives of $v^{2}$ with respect 
to $\mu^{2}$, $\mu_{3}^{2}$, $m_{\tilde{Q}}^{2}$, $m_{\tilde{t}_R}^{2}$ 
and $m_{S}^{2}$ be less than $\Delta$.

One has, taking into account also the variation of $\tan{\beta}$,
\begin{equation}
  \lambda^{2} \delta v^{2}
  = \delta\mu_{3}^{2} \left(t+t^{-1}\right) 
    -\delta\mu_{1}^{2} (2F_{1}-F_{2}) - \delta\mu_{2}^{2} F_{2},
\label{eq:sens}
\end{equation}
where
\begin{equation}
  F_{1} = 1 + \frac{(t-t^{-1})^{2}}{4} 
    \left( \frac{\lambda^{2}v^{2}}{m_{H^{\pm}}^{2}}+1 \right),
\qquad
  F_{2} = 1 + \frac{t^{2}-1}{2} 
    \left( \frac{\lambda^{2}v^{2}}{m_{H^{\pm}}^{2}}+1 \right).
\label{eq:F1-F2}
\end{equation}
Here, we have used the approximation $\mu \simeq m$, $\mu_3^2 \simeq Bm$ 
and $F \simeq 1$, valid for small $s$.

\ From the variation on $\mu_{3}$, upon use of Eq.~(\ref{eq:mA}), 
one obtains, taking $\lambda=2$ and normalizing to a fine-tuning 
$\Delta^{-1} = 20\%$,
\begin{equation}
  m_{A} \simlt 800~\text{GeV}\, (\Delta/5)^{1/2}.
\end{equation}
This therefore sets, through (\ref{eq:mHpm},~\ref{eq:mu-bound}), the 
limits on the charged Higgs-boson and chargino masses (for $\Delta=5$):%
\footnote{The direct naturalness limit on $\mu$ from its contribution 
to $\mu_{1}^{2}$ and $\mu_{2}^{2}$ is less stringent.}
\begin{equation}
  m_{H^{\pm}} \simlt 700~\text{GeV},
\qquad
  \mu \simlt \cos{\beta}\, m_{H^{\pm}}.
\label{eq:mH-bound}
\end{equation}

The stop masses, $m_{\tilde{Q}}$ and $m_{\tilde{t}_{R}}$, and the soft 
breaking mass for the scalar $S$, $m_{S}$, affect $\mathbf{m}_{1}$ and 
$\mathbf{m}_{2}$ through the one loop RGEs
\begin{align}
  \frac{d\mathbf{m}_{1}^{2}}{dt} 
    &= \lambda^{2}\frac{m_{S}^{2}}{8\pi^{2}} + \cdots,
\\
  \frac{d\mathbf{m}_{2}^{2}}{dt} 
    &= \lambda^{2}\frac{m_{S}^{2}}{8\pi^{2}} + \frac{3}{8\pi^{2}}
         \lambda_{t}^{2}(m_{\tilde{Q}}^{2}+m_{\tilde{t}_{R}}^{2}) + \cdots.
\end{align}
There is no particular problem in integrating these equations up to 
the messenger scale $\Lambda_{\rm mess}$ to obtain the standard stop 
contribution
\begin{equation}
  \delta\mathbf{m}_{2}^{2} 
    \simeq -\frac{3}{8\pi^{2}} \lambda_{t}^{2} 
      (m_{\tilde{Q}}^{2}+m_{\tilde{t}_{R}}^{2}) 
      \ln\frac{\Lambda_{\rm mess}}{1~\text{TeV}},
\end{equation}
with an RGE improvement needed if $\Lambda_{\rm mess}$ gets far above 
$100~{\rm TeV}$ or so.  From the sensitivity of $v^{2}$ to $\mu_{2}^{2}$ 
in Eq.~(\ref{eq:sens}), taking $\Lambda_{mess}=100~{\rm TeV}$ as 
a reference point, one obtains
\begin{equation}
  m_{\tilde{t}_{R}},m_{\tilde{Q}} \simlt 
    1.3~\text{TeV} \sin\beta\, (\Delta/5)^{1/2}/\sqrt{F_{2}}.
\end{equation}
The contributions of $m_{S}^2$ to $\mathbf{m}_{i}^{2}$ are, on the other 
hand, not equally well defined due to a rapid increase of $\lambda$ with 
energy.  By integrating the RGEs up to $\Lambda \approx 10~{\rm TeV}$, 
where perturbation theory is still valid, one obtains
\begin{equation}
  \delta\mathbf{m}_{i}^{2}(\Lambda)
  \simeq -\frac{m_{S}^{2}}{4} 
    \ln\left(1+\frac{\lambda_{0}^{2}}{2\pi^{2}}\ln\frac{\Lambda}{500}\right)
  \approx -0.3\, m_{S}^{2},
\end{equation}
where $\lambda_0$ represents the coupling $\lambda$ at $\Lambda$, and, 
consequently,
\begin{equation}
  m_{S} \simlt 1~\text{TeV}\, (\Delta/5)^{1/2}/\sqrt{F_{1}}.
\label{eq:mS-bound}
\end{equation}
Finally, it is of interest to know the naturalness bound on the gluino 
mass $m_{\tilde{g}}$, which contributes to $\mathbf{m}_{2}^{2}$ via 
a two loop effect.  Similarly to the bound on the stop masses, again 
for $\Lambda_{\rm mess}=100~{\rm TeV}$, one obtains
\begin{equation}
  m_{\tilde{g}} \simlt 2.8~\text{TeV} \sin\beta\, 
    (\Delta/5)^{1/2}/\sqrt{F_{2}}.
\end{equation}
Note that all these bounds are proportional to $\lambda$, which is taken 
to be equal to $2$.

\begin{figure}[ptb]
\begin{center}
  \includegraphics[width=10.0cm]{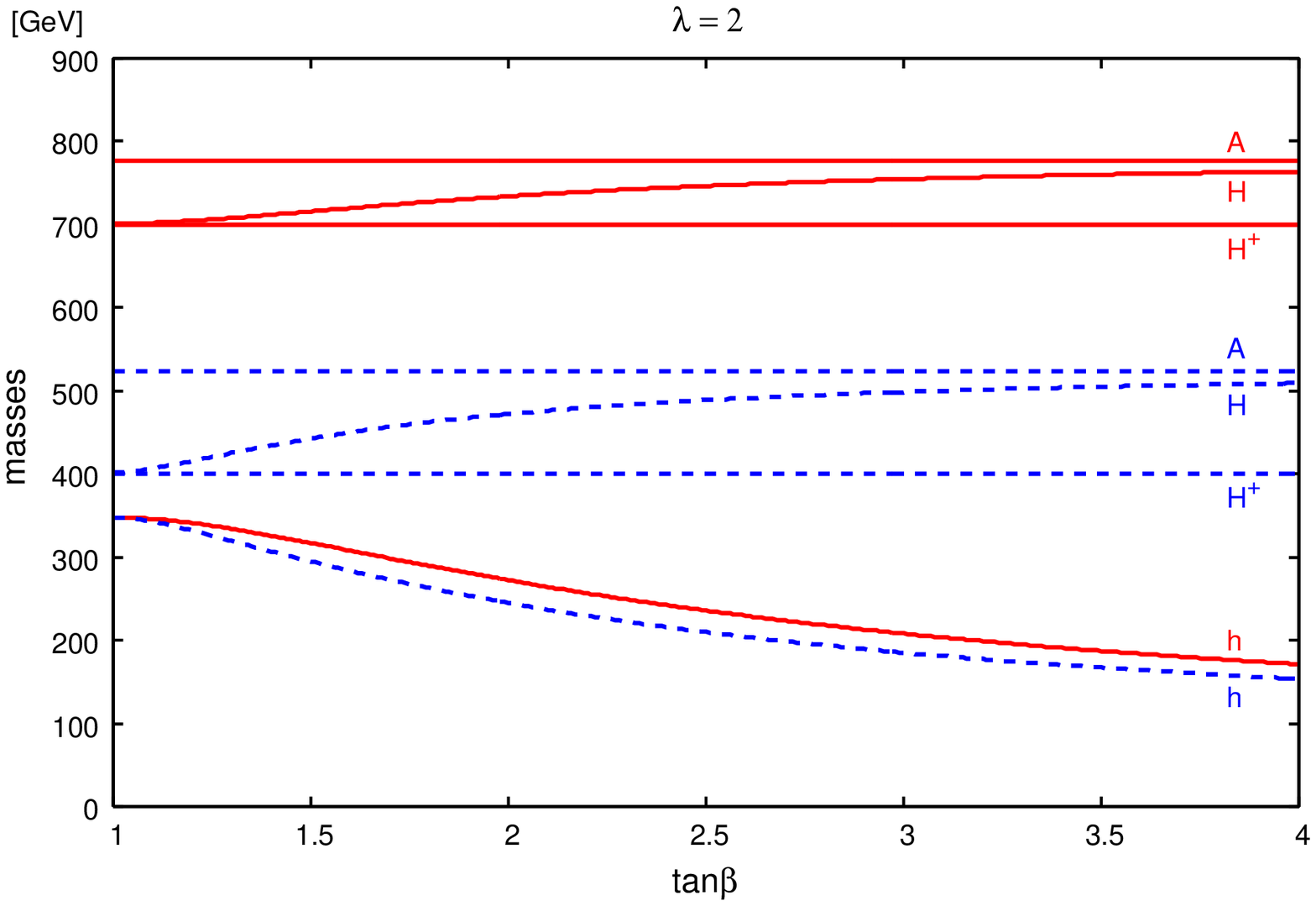}
\end{center}
\vspace{0.1cm}
\begin{center}
  \includegraphics[width=10.0cm]{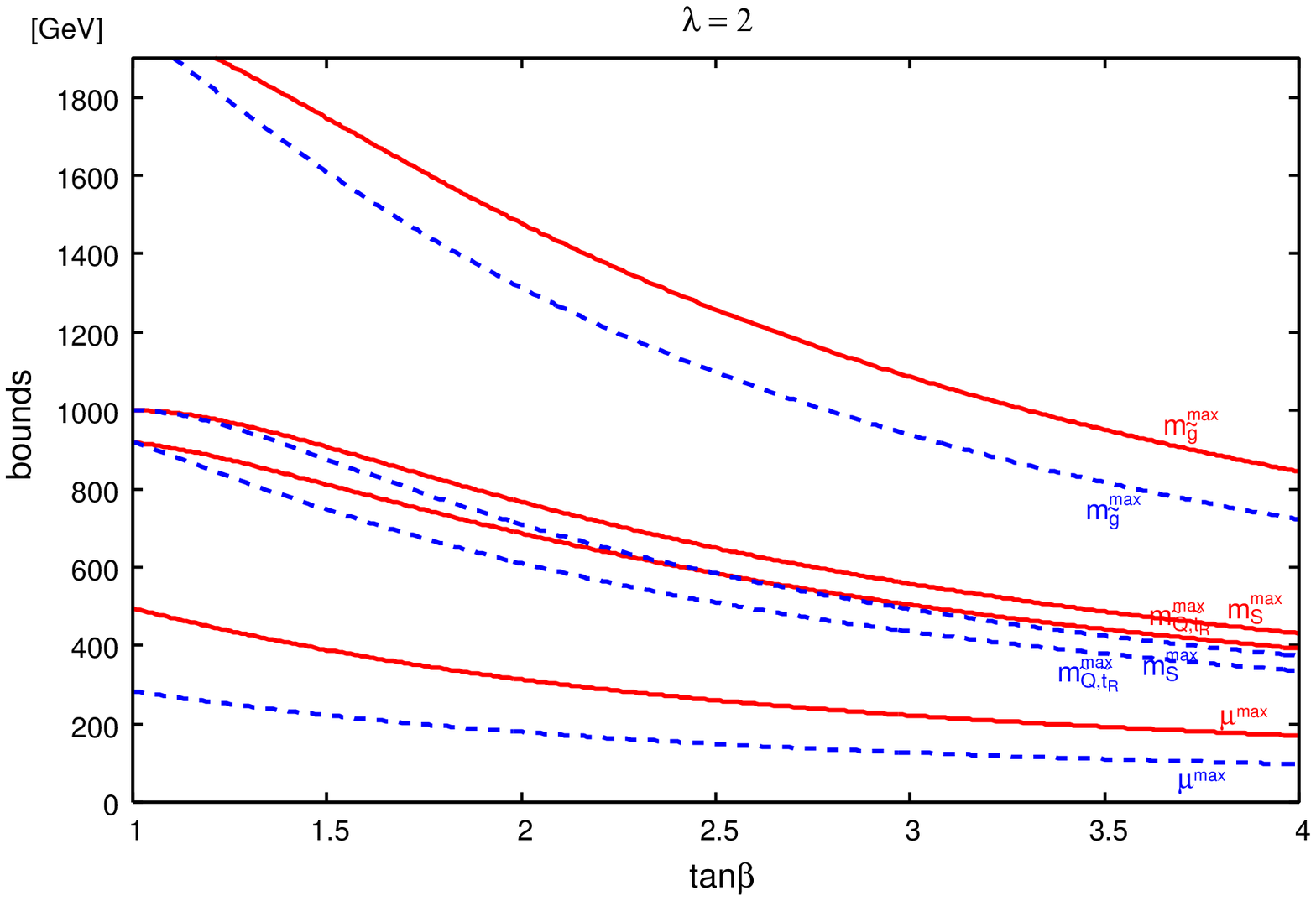}
\end{center}
\caption{The spectrum of the Higgs bosons (top) and the naturalness 
 bounds on parameters (bottom) as functions of $\tan\beta$, for two 
 cases of $m_{H^{\pm}}=700~{\rm GeV}$ (solid lines) and $400~{\rm GeV}$ 
 (dashed lines).  In the top panel, four lines represent $m_{A}$, 
 $m_{H}$, $m_{H^{\pm}}$ and $m_{h}$ from above, for each case 
 of $m_{H^{\pm}}=700, 400~{\rm GeV}$.  In the bottom panel, 
 four lines represent $m_{\tilde{g}}^{\max}$, $m_{S}^{\max}$, 
 $m_{\tilde{Q},\tilde{t}_{R}}^{\max}$ and $\mu^{\max}$ from above, 
 for each case of $m_{H^{\pm}}=700, 400~{\rm GeV}$.}
\label{fig:spectrum}
\end{figure}

Figure~\ref{fig:spectrum} summarizes the knowledge of the spectrum 
for different values of $\tan{\beta}$ and for two reference values 
of $m_{H^{\pm}} = 700, 400~{\rm GeV}$.  The former corresponds 
to the highest value compatible with the naturalness bound of 
Eq.~(\ref{eq:mH-bound}), and both values are consistent with the 
constraint from $b\rightarrow s \gamma$ without a destructive 
contribution from a stop-chargino loop~\cite{Gambino:2001ew}.
The top and bottom panels show, respectively, the values of the Higgs 
boson masses and the upper bounds from naturalness on $\mu$, $m_{S}$, 
$m_{\tilde{Q},\tilde{t}_R}$, and $m_{\tilde{g}}$, allowed by a $20\%$ 
tuning.  The bound on the $S$-fermion mass $M$ is not explicitly given, 
since it depends, via $F > 1/2$, on many other parameters.

We finally remark that the naturalness bounds derived from variations of 
$\tan\beta$ alone are less restrictive than the above bounds derived from 
the variations of $v^{2}$.

\section{LHC Phenomenology}
\label{sec:LHC}

How will LHC probe $\lambda$SUSY --- the theory defined by (\ref{eq:superpot}) 
and (\ref{eq:pot})?  Clearly we must consider signals that result from the 
spectrum and mixings of the Higgs and Higgsinos, paying particular attention 
to the consequences of a large value of $\lambda$, and to differences with 
the MSSM.  A general and detailed treatment is far beyond the scope of 
this paper; we will be content with discussing a few illustrative examples, 
especially as they relate to a collider probe of DM.

Throughout we assume that the $S$ scalar is heavy and its mixing with 
the doublet scalars can be ignored.  In this case, the parameters relevant 
for the doublet scalars are $\mu^2_{1,2,3}$ and $\lambda$, which are 
reduced to 3 by the minimization condition for the VEV $v$.  Including 
both neutral and charged Higgsinos, in the limit that the gauginos are 
sufficiently heavy, there are an additional 2 parameters $\mu$ and $M$, 
giving a total of 5 free parameters.  It might appear that there are 
2 extra parameters in this sector compared to the MSSM, $\lambda$ and 
$M$, but this is not the case --- in the MSSM one needs an enhanced 
quartic coupling that brings in the top squark sector.  Hence there is 
a single extra parameter, $M$, that describes the mass and mixings of 
an extra neutralino, so that the theory is highly constrained.  Furthermore, 
the addition of the singlino allows for the possibility of Higgsino DM.

The four doublet scalars themselves show important differences with 
the MSSM.  First and foremost the lighter Higgs $h$ is much heavier 
than in the MSSM. Second, the ordering of the spectrum is fixed: $h$, 
$H^\pm$, $H$ with $A$ heaviest.%
\footnote{Parameter choices allow $H^\pm$ to be lighter than $h$, 
but this must be consistent with the limit on $m_{H^\pm}$ from 
$b \rightarrow s \gamma$.}  
In contrast, for the MSSM the ordering of $H^\pm$, $A$ and $H$ is 
not fixed, except $A$ is lighter than $H$.  Finally, $\tan\beta$ less 
than about 3 is excluded in the MSSM, while it is strongly preferred 
in $\lambda$SUSY, leading to important changes in the various decay 
rates for these scalars.  The three neutral scalars $h,H,A$ are each 
copiously produced at LHC from gluon-gluon fusion with cross sections 
in the range $0.1$~--~$10~{\rm pb}$, decreasing with mass.  While 
these rates can be computed in terms of the Higgs masses and mixings, 
$h$, $H$ and $A$ may also be produced in cascade decays of top and 
bottom squarks (and from other squarks, if light enough to be produced) 
so that we will not consider the rates themselves for determining 
the 5 free parameters, but will turn to the masses and widths of 
the three neutral scalars.

With a branching ratio of about $10^{-3}$ in much of parameter space, 
$h$ and $H$ will be visible in ``gold-plated'' events where the scalar 
decays to $ZZ$, and each $Z$ decays to $l \bar{l}$, where $l$ is an 
electron or muon.  This will allow accurate measurements of $m_h$ and 
$m_H$.   Similarly, the decay $A \rightarrow Zh$ will allow a measurement 
of $m_A$ from events of the form $l \bar{l} l \bar{l} jj$ where each 
$l \bar{l}$ reconstructs to a $Z$, as does the two jet system $jj$.%
\footnote{As $m_H$ increases and approaches its naturalness limit of 
$700~{\rm GeV}$ the decoupling regime is reached, so that the decays 
$H \rightarrow ZZ$ and $A \rightarrow Zh$ become suppressed relative 
to the $t\bar{t}$ mode.  Since the production cross section is also 
reduced, mass measurements from the gold-plated decays may be difficult 
in this region.  However, this region is not the most natural expectation. 
Note that while the couplings of either $h$ or $H$ to vector boson pairs 
vanish in $\lambda$SUSY as $\tan\beta \rightarrow 1$, this suppression 
becomes rapidly ineffective as $\tan\beta$ starts deviating from 1.}
The measurements of $m_{h,H,A}$, together with the constraint from 
the electroweak VEV, will allow all 4 parameters of the Higgs sector 
to be determined.  This includes both the quartic coupling $\lambda$, 
so that one can evaluate the scale at which non-perturbative physics 
sets in, and $\tan\beta = \mu_1/\mu_2$.  Unlike the MSSM, the ratio 
of VEVs can be determined relatively easily.  Notice that $m_{h,H,A}$, 
and hence the extraction of the 4 fundamental parameters, do not 
depend on any assumption about the superpartner spectrum.

To extract $\mu$ and $M$, we specialize to the case that the gauginos 
are sufficiently heavier than the Higgsinos that we can ignore the 
Higgsino/gaugino mixing.  In practice the accuracy with which the 
parameters can be extracted will depend on the size of the gaugino 
components of the lighter neutralinos and charginos.  Over much of 
the parameter space, the decay of $H$, $A$ and $h$ to Higgsino pairs 
is kinematically allowed.  Since it is induced at tree-level by 
the large coupling $\lambda$, it will be competitive with the $WW$ 
mode, generically having a significant effect on the total widths 
$\Gamma_{H,A,h}$.  Note that the $WW,ZZ$ and $t\bar{t}$ modes depend 
on the Higgs mixing parameters and on $\tan\beta$ (needed for example 
to compute the top Yukawa coupling).  However, they can be accurately 
computed, since these parameters will have been determined by the 
mass measurements.  The widths to Higgsinos depend on both the Higgs 
mixings, already determined by the Higgs masses, and by the Higgsino 
mixings and masses, which are determined by $\mu$ and $M$.  Can the 
total widths $\Gamma_{H,A,h}$ be measured sufficiently accurately, by 
reconstructing the invariant mass from gold-plated events, to extract 
$\mu$ and $M$?  For $A \rightarrow Zh$ there will be insufficient 
rate to rely on the $6l$ events, and the $4l jj$ events are unlikely 
to be reconstructed with sufficient accuracy.  Hence we consider 
the possibility of measuring $\mu$ and $M$ via the contributions 
of the Higgsino pair mode to $\Gamma_H$ and $\Gamma_h$, reconstructing 
these total widths using the $4l$ final states.

\begin{figure}[ptb]
\begin{center}
  \includegraphics[width=8.5cm]{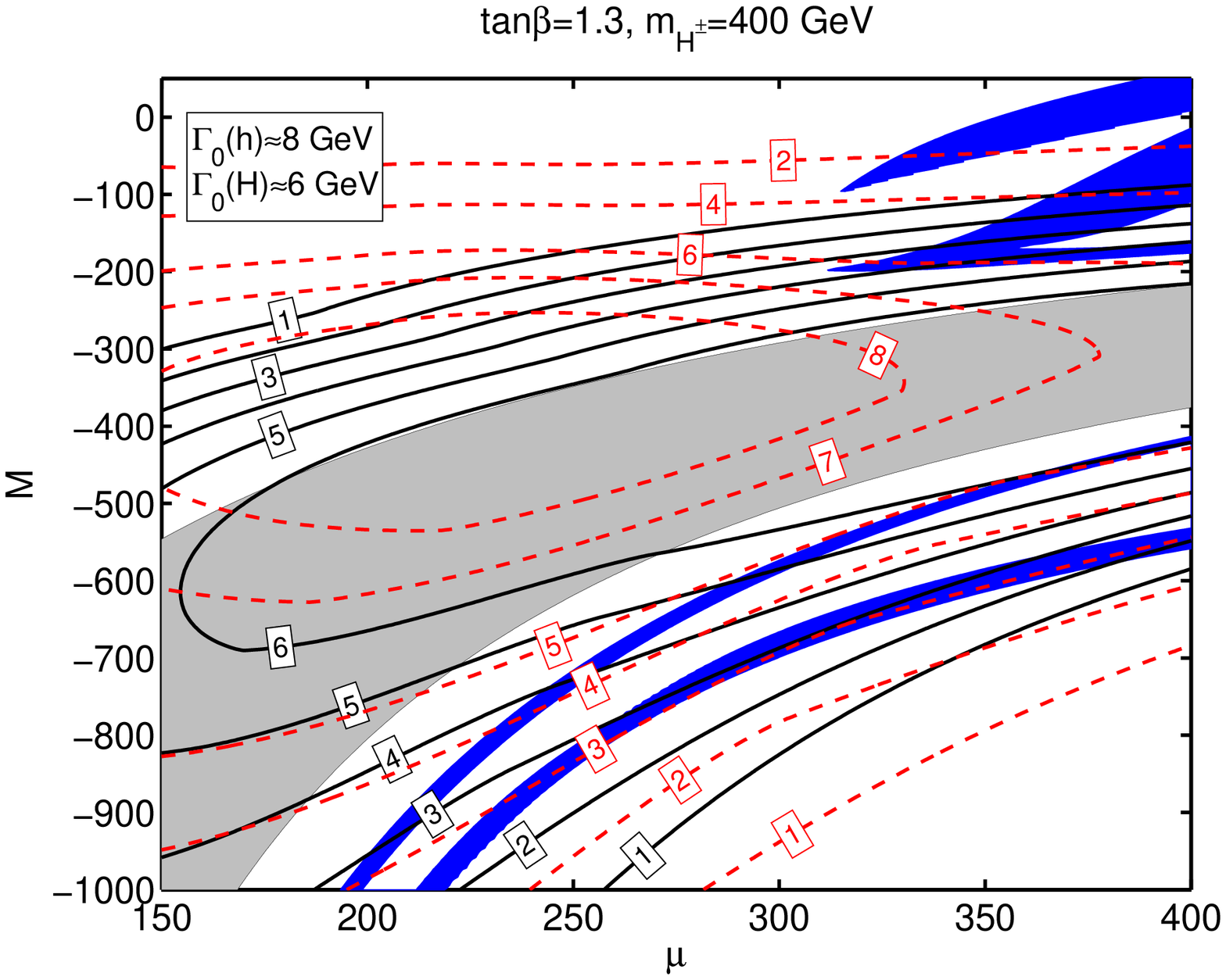}
  \includegraphics[width=8.5cm]{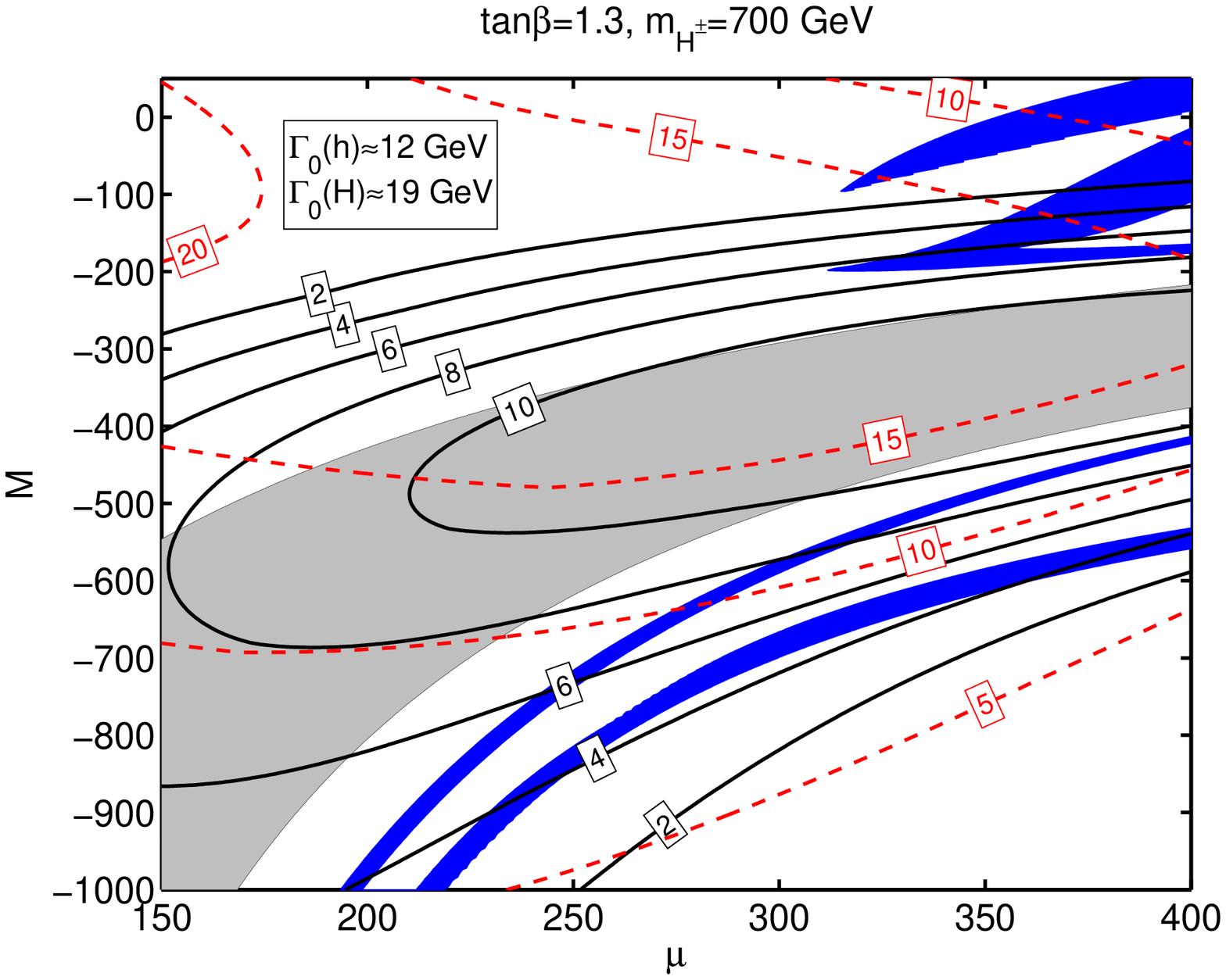}
  \includegraphics[width=8.5cm]{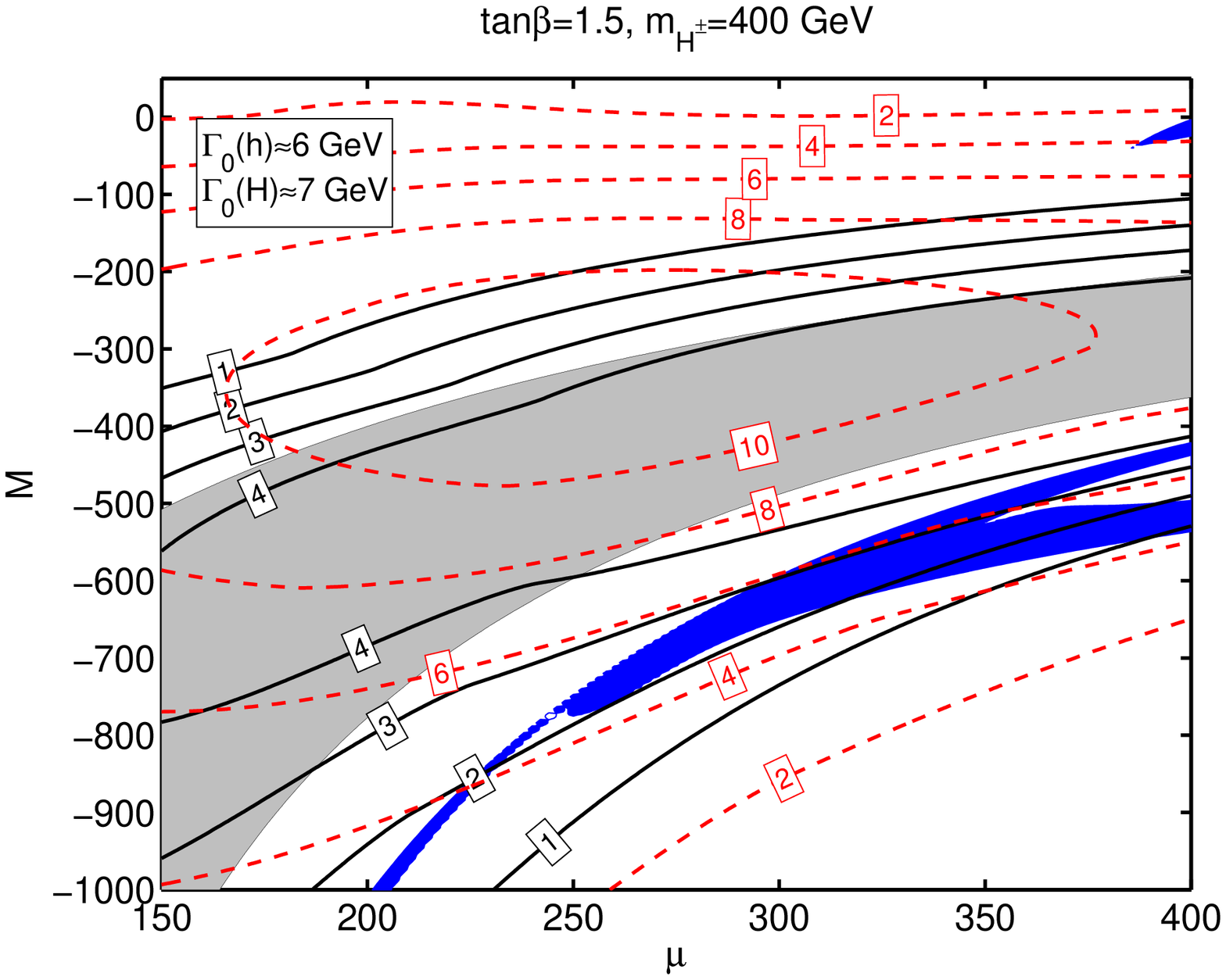}
  \includegraphics[width=8.5cm]{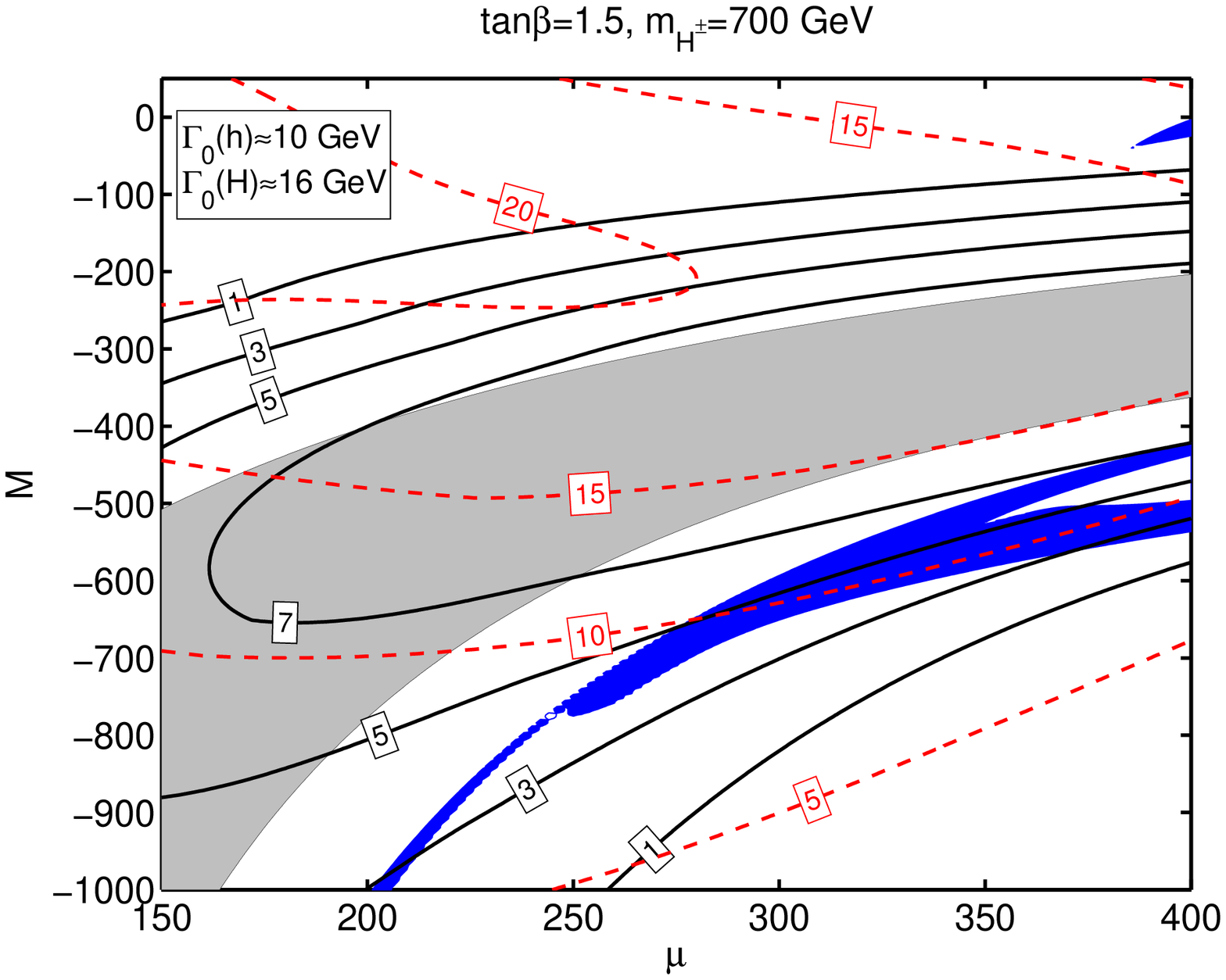}
\end{center}
\caption{Contours for the partial widths $H \rightarrow {\rm Higgsinos}$ 
 (dashed red curves) and $h \rightarrow {\rm Higgsinos}$ (solid black 
 curves), labeled in GeV, in the $\mu$-$M$ plane for $\lambda = 2$.  The 
 four panels are for $\tan\beta = 1.3,1.5$ and $m_{H^\pm} = 400,700~{\rm GeV}$. 
 For each of the four panels we give the dominant non-Higgsino widths, 
 $\Gamma_0$, for comparison.  The gray shaded regions have $m_\chi < m_Z/2$, 
 while the dark (blue) shaded region corresponds to $0.09 \simlt \Omega_{\chi} 
 h^{2} \simlt 0.13$.}
\label{fig:width-1}
\end{figure}

The partial widths to Higgsinos for $H$ and $h$ are shown as contours 
in Fig.~\ref{fig:width-1} for $\lambda=2$, $\tan\beta = 1.3,1.5$ and 
$m_{H^\pm} = 400, 700~{\rm GeV}$ in the $\mu$-$M$ plane.
For each of the four panels we give the dominant non-Higgsino widths 
for comparison: $\Gamma_0(h) = \Gamma(h \rightarrow WW + ZZ)$ and 
$\Gamma_0(H) = \Gamma(H \rightarrow WW + ZZ + t\bar{t} + hh)$.  Consider 
for example the panel with $\tan\beta = 1.3$ and $m_{H^\pm} = 400~{\rm 
GeV}$, where $\Gamma_0(h) \approx 8~{\rm GeV}$ and $\Gamma_0(H) \approx 
6~{\rm GeV}$.  From inspection of the contours, we find that the Higgsino 
widths for both $h$ and $H$ are in the range of $(10\!\sim\!100)\%$ of 
the corresponding non-Higgsino widths, allowing a determination of 
$\mu$ and $M$ over much of the parameter space.  The accuracy of the 
determination will depend on the location in the $\mu$-$M$ plane. 
For much of the DM region, shown shaded in Fig.~\ref{fig:width-1}, the 
percentage changes in the widths induced by the Higgsino mode is roughly 
in the region of $30\%$.  Increasing $m_{H^\pm}$ to $700~{\rm GeV}$ 
leads to similar percentage changes in the widths, so again measurements 
of $\mu$ and $M$ are possible.  The number of gold-plated events from 
$H$ decay decreases as both $m_H$ and $\tan\beta$ are increased, making 
the measurement more difficult as the regions disfavored by naturalness 
and EWPT are approached.  It will, of course, be highly significant 
if measurements at the LHC indicate values of $\mu$ and $M$ in the DM 
region, determining whether the DM Higgsino is mainly singlino or has 
comparable doublet and singlet components, and allowing predictions for 
the direct detection rate and the LSP mass.  As $\tan\beta$ is increased 
from $1.3$ to $1.5$ the signals in $\Gamma_{h,H}$ are comparable, but 
the dominantly singlino region is essentially absent, so it will be 
an easier task to verify consistency with the remaining DM region. 

\begin{figure}[ptb]
\begin{center}
  \includegraphics[width=8.5cm]{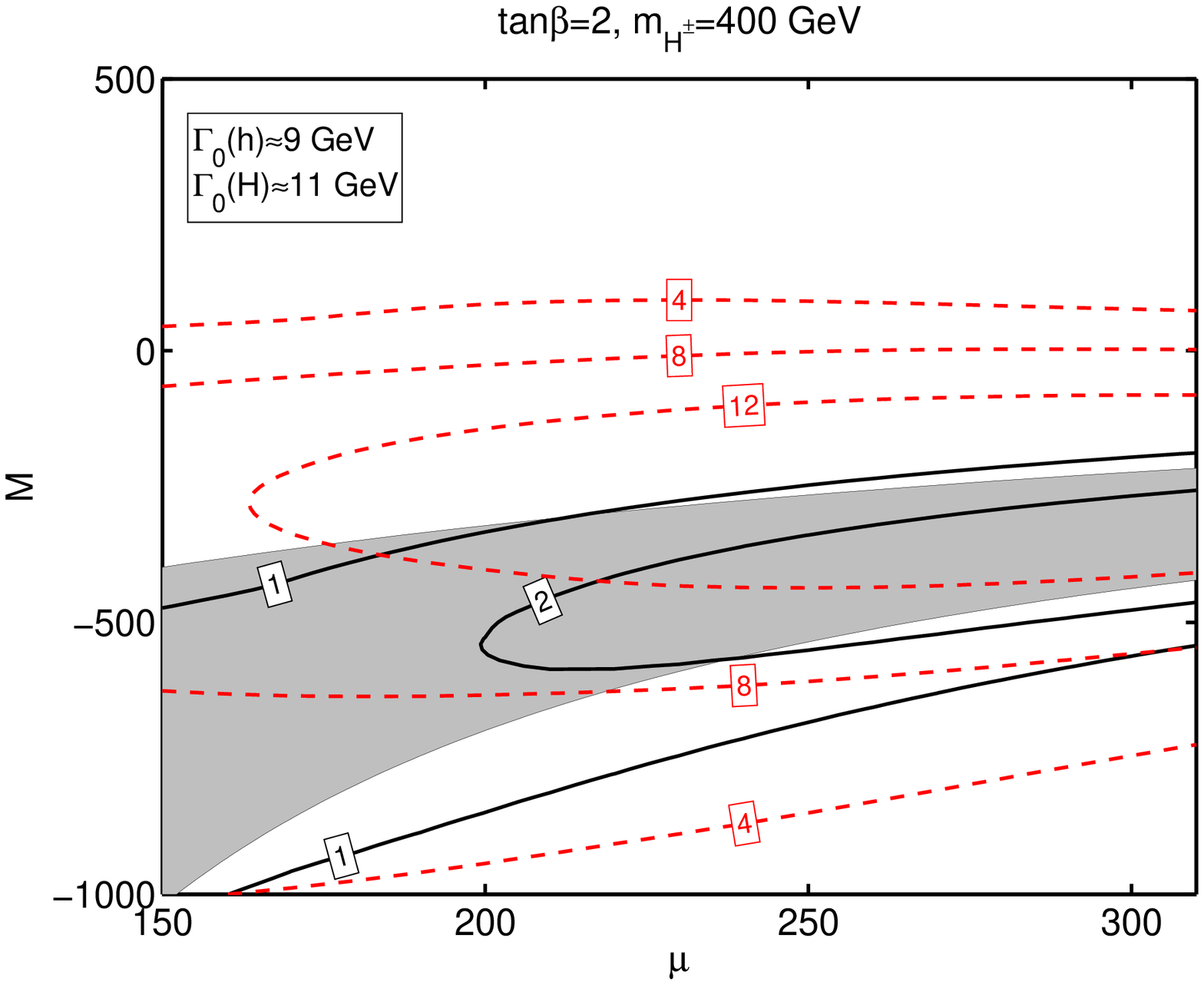}
  \includegraphics[width=8.5cm]{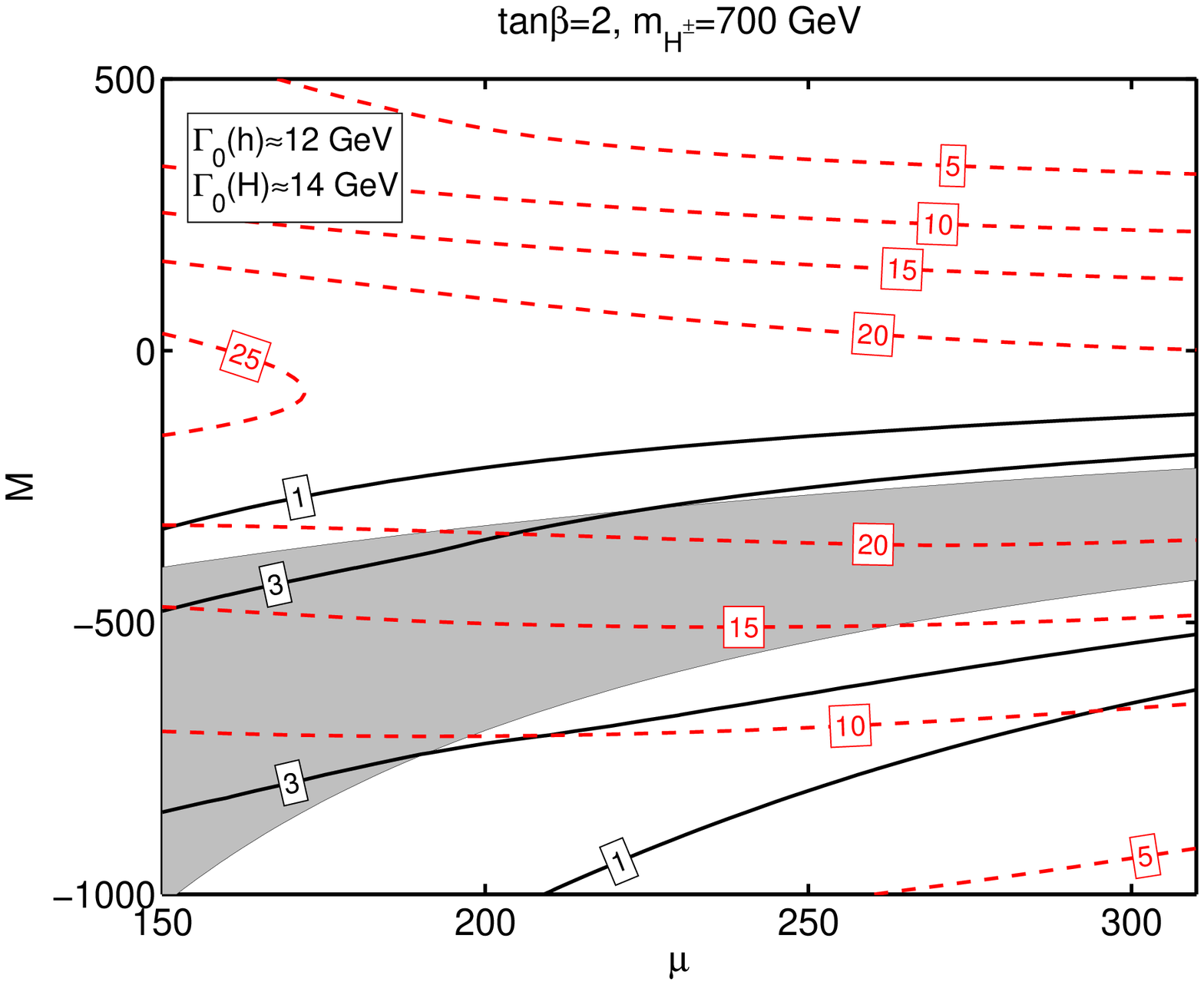}
  \includegraphics[width=8.5cm]{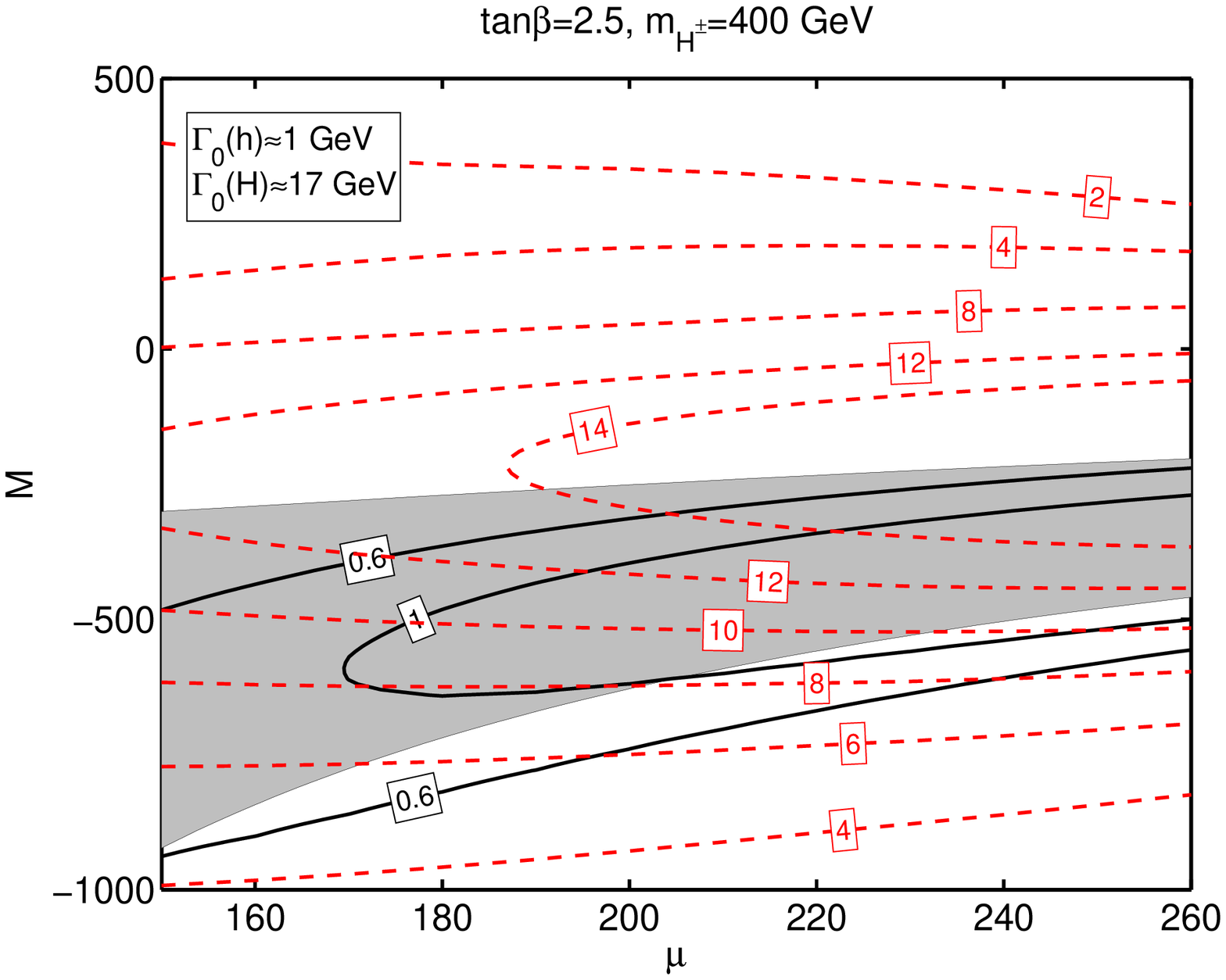}
  \includegraphics[width=8.5cm]{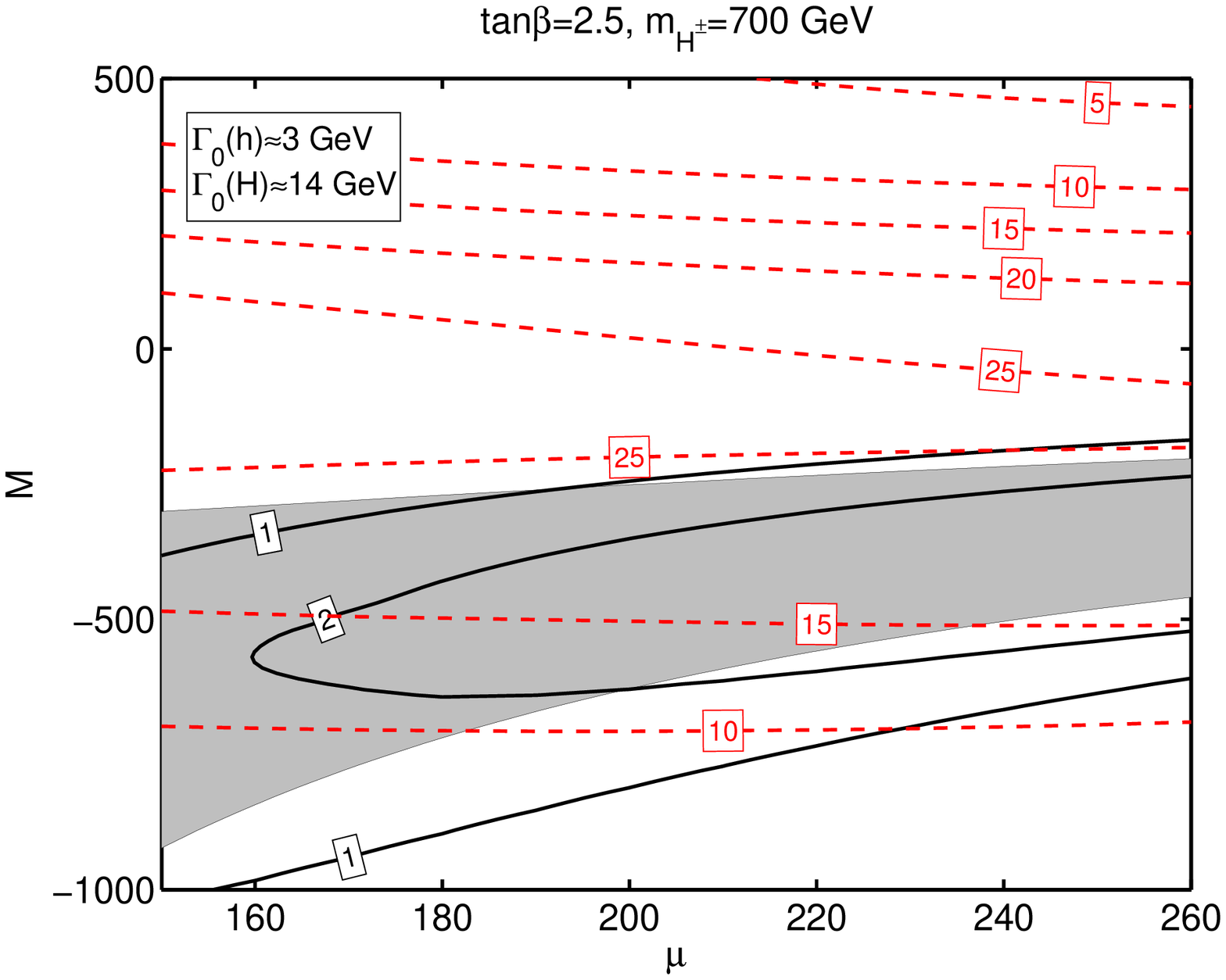}
\end{center}
\caption{As for Fig.~\ref{fig:width-1}, except for $\tan\beta = 2, 2.5$. 
 There are no regions giving sufficient thermal relic Higgsino DM.  Note 
 that a larger region of $M$ is shown than in Fig.~\ref{fig:width-1}, 
 but smaller regions in $\mu$, corresponding to the more stringent 
 naturalness constraints at larger $\tan\beta$.}
\label{fig:width-2}
\end{figure}

In Fig.~\ref{fig:width-2} we show the contours for $\Gamma(H,h \rightarrow 
{\rm Higgsinos})$ for larger values of $\tan\beta$ that do not allow 
thermal relic Higgsinos to be DM. 
Nevertheless, the change in the total widths induced by the Higgsino 
contributions is again a significant percentage, so that $\mu$ and $M$ 
can be measured.

Determining parameters from studying the decay modes of Higgsinos may 
also be possible, but it looks to be very difficult.  Consider pair 
production of top squarks, followed by decays to $t+ {\rm Higgsino}$. 
The heavier neutral Higgsinos cascade to the LSP and a Higgs boson, so 
that the decay chain is $\tilde{t} \rightarrow t \tilde{\chi}_{2,3}$ 
followed by $\tilde{\chi}_{2,3} \rightarrow \tilde{\chi}_1 H$ or 
$\tilde{\chi}_{2,3} \rightarrow \tilde{\chi}_1 h$, leading to signals 
$t \bar{t} l \bar{l} l \bar{l}$ with the four leptons reconstructing to 
$m_H$ or $m_h$.  The ratio of these two types of events is another probe 
of the parameter space, since it is independent of the mass and production 
rate of the top squarks.  However, this could be contaminated by events 
with bottom squarks decaying to top quarks and the charged Higgsino, 
which itself cascades to $H$ or $h$.

\section{Overview and Conclusions}
\label{sec:OC}

One year from now the Large Hadron Collider will start directly exploring 
for the first time the energy range well above the Fermi scale.  Many 
speculations of the last three decades on ElectroWeak Symmetry Breaking 
(EWSB) will be rendered irrelevant, while some, perhaps, may emerge as 
part of physical reality.  Therefore, the time is right to briefly reconsider 
the status of these speculations, including the one presented in this work. 

There are two main sets of considerations that have played a dominant role 
to orient ideas and specific work on the subject.  One rotates around the 
naturalness problem of the Fermi scale.  Why is the Higgs boson light relative 
to any mass scale appearing in the UV completion of the SM, whatever it may be? 
More quantitatively, as illustrated by all known examples that attack the 
problem, why is the Higgs mass not at least as large as the SM contribution 
with a cutoff scale where new physics sets in, $\Lambda_{\rm NP}$?  The 
second set of considerations relates to the significance of the EWPT to 
the EWSB problem.  If new physics occurs at $\Lambda_{\rm NP}$, would 
it not be manifest, even indirectly, in EWPT?

In principle these two kinds of considerations do not conflict with each 
other: they might have actually merged into a coherent picture for the 
physics that underlies EWSB.  So far, it is fair to say that this has 
not happened.  On the contrary, the impressive success of the SM in 
providing an accurate perturbative description of the EWPT casts doubts 
on the physical relevance of many ideas put forward to attack the 
naturalness problem of the Fermi scale.  This issue, the ``little 
hierarchy problem,'' is a quantitative one~\cite{Barbieri:2000gf}: 
for $\Lambda_{\rm NP}$ as low as implied in various attempts to attack 
the naturalness problem, why is the new physics not manifest in the EWPT? 
At first sight the supersymmetric attempts seem exempt from this difficulty, 
as they cope easily with the EWPT.  In their simplest versions, however, 
they also have a problem accommodating the direct bound on the Higgs mass. 
To satisfy the bound apparently requires either a lack of naturalness 
or a complication of the theory.

It would neither be possible nor appropriate to recall here all the 
attempts to get around this difficulty.  Nevertheless one aspect is worth 
emphasizing, as trivial as it may be.  While ignoring the naturalness 
problem is not in contradiction with experiment, it has however a striking 
practical consequence, other than being theoretically unsatisfactory: it 
weakens the case for finding new physics at the LHC.  Needless to say, 
what will be found at the LHC does not depend on our choice of the 
physically relevant problems.  However, insisting on the naturalness 
problem, as we do in this paper, has at least the advantage that 
the proposals to solve it will be scrutinized when the LHC will be 
successfully turned on.

With this motivation in mind, the framework analyzed in this paper 
originates from the observation that the little hierarchy problem might 
have (part of) its source in a misinterpretation of the EWPT, especially 
related to the inferred strong bound on the Higgs mass.  Maybe the results 
of the EWPT, although consistent with the SM, do hide after all some new 
physics.  This might be especially the case if the corresponding new theory 
still allows a successful perturbative description of the EWPT in a large 
portion of its parameter space.  At first sight, this does not seem to 
be of relevance to supersymmetry, since, as observed above, standard 
supersymmetry has no special problem with the EWPT.  However supersymmetry 
in its minimal version has a theoretical bound on the Higgs mass which is 
even stronger than the indirect experimental bound from the EWPT, causing 
its own naturalness problems.  Hence the proposal that we make here of 
relaxing as much as possible the upper bound on the Higgs mass by going 
to $\lambda$SUSY, while still retaining the successful perturbative 
description of the EWPT.  In our view, this ``bottom-up'' approach 
to the little hierarchy problem deserves attention.

We can summarize this bottom-up approach to supersymmetry as follows: 
the Higgs sector is pushed up to the (200~--~700)~GeV range, the 
superpartners may then be heavier than $700~{\rm GeV}$, while most 
of the parameter space with successful EWSB naturally yields the 
observed values for the $S$ and $T$ parameters.

Needless to say $\lambda$SUSY has two related prices to pay, which have 
certainly restrained most people from considering it seriously up to now. 
The first is that it is not a UV-complete theory in the same sense that 
the MSSM or the standard NMSSM are: at an energy scale above $10~{\rm TeV}$ 
or so, some change of regime of the theory should intervene that prevents 
the coupling $\lambda$ from exploding.  This is not uncommon in most 
attempts to address the little hierarchy problem.  The other price is 
that $\lambda$SUSY is not manifestly consistent with gauge coupling 
unification.  To determine this would require specifying the form 
of the necessary change of regime at high energy; it is nevertheless 
remarkable that a successful solution already exists in the 
literature~\cite{Harnik:2003rs,Chang:2004db,Birkedal:2004zx}.

\section*{Acknowledgments}

The work of R.B. and V.S.R. was supported by the EU under RTN contract 
MRTN-CT-2004-503369.  R.B. was also supported in part by MIUR and by a 
Humbolt Research Award.  R.B. thanks the Institute for Theoretical Physics 
of the Heidelberg University for hospitality while this work was completed. 
The work of L.J.H. and Y.N. was supported in part by the US Department of 
Energy under Contracts DE-AC03-76SF00098.  L.J.H. was also supported in part 
by the National Science Foundation under grants PHY-00-98840 and PHY-04-57315, 
and Y.N. by the National Science Foundation under grant PHY-0403380, by 
a DOE Outstanding Junior Investigator award, and by an Alfred P. Sloan 
Research Fellowship.

\appendix

\section{One-loop Contributions of New Particles to {\boldmath $S$} 
and {\boldmath $T$}}
\label{app:A}

Recall that the $T$ and $S$ parameters are given by
\begin{align}
  T &= \frac{A_{33}(0)-A_{WW}(0)}{\alpha_{\rm em} M_{W}^{2}},
\\
  S &= \frac{4s_{\text{w}}c_{\text{w}}}{\alpha_{\rm em}}F_{30}(M_{Z}^{2})
    \approx \frac{4s_{\text{w}}c_{\text{w}}}{\alpha_{\rm em}}F_{30}(0),
\end{align}
where $A_{ij}$ and $F_{ij}$ are extracted from the gauge-boson pair vacuum 
polarization amplitudes:
\begin{equation}
  \Pi_{\mu\nu}^{ij}(q) = -ig_{\mu\nu}[A^{ij}(0)+q^{2}F^{ij}(q^{2})]
    + q_{\mu}q_{\nu}\text{terms},
\end{equation}
with $i,j=W,\gamma,Z$, or $i,j=0,3$ for $B_\mu$ and $W_{3,\mu}$, 
$\alpha_{\rm em}$ is the structure constant, and $s_{\text{w}} \equiv 
\sin\theta_W$ and $c_{\text{w}} \equiv \cos\theta_W$ with $\theta_W$ 
the Weinberg angle.

We record here one-loop expressions for $A(0)$~\cite{Barbieri:1983wy} and 
$F(0)$ produced by bosons and fermions coupled to a generic gauge boson 
$W_{\mu}$ with unit strength.  For a \textit{boson} loop with internal 
masses $m_{1}$ and $m_{2}$ and coupling $iW_{\mu}\phi_{1}^{\ast} 
\overleftrightarrow{\partial}_{\!\!\mu} \phi_{2}$ we have 
\begin{align}
  A(0) = 
  & \frac{1}{16\pi^{2}} \biggl[ \frac{m_{1}^{2}+m_{2}^{2}}{2}
    - \frac{m_{1}^{2}m_{2}^{2}}{m_{1}^{2}-m_{2}^{2}}
      \ln\frac{m_{1}^{2}}{m_{2}^{2}} \biggr] 
    \equiv 2 \alpha_{\rm em} v^{2} A(m_{1},m_{2}),
\\
  F(0) = 
  & \frac{1}{96\pi^{2}} \biggl[ -\ln\frac{\Lambda^{4}}{m_{1}^{2}m_{2}^{2}}
    + \frac{4m_{1}^{2}m_{2}^{2}}{(m_{1}^{2}-m_{2}^{2})^{2}}
\nonumber\\
  & + \frac{m_{1}^{6}+m_{2}^{6}-3m_{1}^{2}m_{2}^{2}(m_{1}^{2}+m_{2}^{2})}
      {(m_{1}^{2}-m_{2}^{2})^{3}} \ln\frac{m_{1}^{2}}{m_{2}^{2}} \biggr]
    \equiv \frac{1}{4\pi} F(m_{1},m_{2}).
\end{align}
There is an additional diagram contributing to $S$ for a Higgs boson, in 
which the gauge boson $W_\mu$ and the Higgs boson $\phi$ propagate in the 
loop, giving $\delta S = m_W^2 G(m_\phi, m_W)$ where
\begin{equation}
  G(m_{1},m_{2}) \equiv \frac{1}{2\pi}
    \left[ \frac{2m_{1}^{2}m_{2}^{2}}{(m_{1}^{2}-m_{2}^{2})^{3}}
      \ln\frac{m_{1}^{2}}{m_{2}^{2}}
    - \frac{m_{1}^{2}+m_{2}^{2}}{(m_{1}^{2}-m_{2}^{2})^{2}} \right].
\end{equation}

For a \textit{fermion} loop with internal masses $m_{1}$ and $m_{2}$ and 
a vector coupling $W_{\mu}\bar{\psi}_{1}\gamma^{\mu}\psi_{2}$, we have
\begin{align}
  A(0) =
  & \frac{1}{16\pi^{2}} \biggl[ (m_{1}-m_{2})^{2}\ln\frac{\Lambda^{4}}
      {m_{1}^{2}m_{2}^{2}} - 2m_{1}m_{2}
\nonumber\\
  & + \frac{2m_{1}m_{2}(m_{1}^{2}+m_{2}^{2})-m_{1}^{4}-m_{2}^{4}}
      {m_{1}^{2}-m_{2}^{2}} \ln\frac{m_{1}^{2}}{m_{2}^{2}} \biggr]
    \equiv 2 \alpha_{\rm em} v^{2} \tilde{A}(m_{1},m_{2}),
\\
  F(0) =
  & \frac{1}{24\pi^{2}} \biggl[ -\ln\frac{\Lambda^{4}}{m_{1}^{2}m_{2}^{2}}
    - \frac{m_{1}m_{2}(3m_{1}^{2}-4m_{1}m_{2}+3m_{2}^{2})}
      {(m_{1}^{2}-m_{2}^{2})^{2}}
\nonumber\\
  & + \frac{m_{1}^{6}+m_{2}^{6}-3m_{1}^{2}m_{2}^{2}(m_{1}^{2}+m_{2}^{2})
      +6m_{1}^{3}m_{2}^{3}}{(m_{1}^{2}-m_{2}^{2})^{3}} 
      \ln\frac{m_{1}^{2}}{m_{2}^{2}} \biggr]
    \equiv \frac{1}{4\pi} \tilde{F}(m_{1},m_{2}).
\end{align}
For an axial coupling, the result can be obtained by letting $m_{1} \rightarrow 
-m_{1}$.  These expressions are valid for both Dirac and Majorana fermions, 
with an extra factor of 2 in the case of identical Majorana fermions.


\end{document}